\documentclass[
twocolumn,
superscriptaddress,
nofootinbib,
 amsmath,amssymb,
 aps,
prb,
]{revtex4-2}

\usepackage{graphicx}
\usepackage{dcolumn}
\usepackage{bm}
\usepackage{xcolor}
\usepackage{hyperref}
\usepackage{multirow}
\usepackage{physics}

\begin{document}

\title{Supervised learning of an interacting 2D hard-core boson model of a weak topological insulator using correlation functions}

\author{Amrita Ghosh}%
 \email{amrita.ghosh.physics@gmail.com}
 \affiliation{Center for Theory and Computation, National Tsing Hua University, Hsinchu 300, Taiwan}
 \affiliation{Physics Division, National Center for Theoretical Sciences, National Taiwan University, Taipei 10617, Taiwan}
\author{Mugdha Sarkar}%
\email{mugdha.sarkar@gmail.com}
\affiliation{Department of Physics, National Taiwan University, Taipei 10617, Taiwan}
\affiliation{Physics Division, National Center for Theoretical Sciences, National Taiwan University, Taipei 10617, Taiwan}

\begin{abstract}
We study a system of hard-core bosons on a two-dimensional periodic honeycomb lattice subjected to an on-site potential with alternating signs along $y$-direction, using machine learning (ML) techniques. The model hosts a rich phase diagram consisting of six different phases including a charge density wave, a superfluid phase and two dimer insulator phases, one of which is also a weak topological insulator with zero Chern number but a non-trivial Berry phase [\href{https://doi.org/10.21468/SciPostPhys.10.3.059}{SciPost Phys.~\textbf{10}, 059 (2021)}]. Using two distinct correlation functions computed via quantum Monte Carlo method, a relatively simple ML model is able to learn information from the various phases simultaneously and accurately predict their phase boundaries. By employing our ML model trained on {the dataset obtained from the Hamiltonian without explicit interactions}, we determine the phase structure of the system in the presence of nearest-neighbor interactions. Additionally, we investigate the robustness of the weak topological insulator phase against interactions by predicting the topological invariant, which is otherwise difficult to obtain.
\end{abstract}

\maketitle

\section{Introduction}\label{sec:Intro}
Since the discovery of neural networks and machine learning (ML) techniques, there has been a rapid growth in interest in the application of these methods to condensed matter physics (for recent reviews, see Ref.~\cite{Dawid:2022fga,Bedolla_2021,RevModPhys.91.045002,doi:10.1080/23746149.2020.1797528}). ML techniques have been successfully applied to predict phase transitions and characterize various phases, using supervised and unsupervised learning on different kinds of input data \cite{PhysRevB.94.195105,doi:10.7566/JPSJ.85.123706,carrasquilla_machine_2017,vanNieuwenburg2017,doi:10.7566/JPSJ.86.063001,PhysRevB.96.184410,PhysRevE.95.062122,https://doi.org/10.48550/arxiv.1707.00663,PhysRevE.97.013306,PhysRevLett.120.176401,PhysRevB.97.134109}. In particular, correlation functions have proven to be highly effective in training ML models to acquire knowledge about different phases \cite{PhysRevB.100.045129,Shiina2020,PhysRevE.102.021302,PhysRevResearch.3.033052,10.21468/SciPostPhys.14.1.005,chung2023deep}. 

Topological insulators (TIs) are an exciting area of research within the field of topological phases of matter, that feature a bulk band gap like an ordinary insulator but also possess topologically-protected gapless surface states \cite{RevModPhys.82.3045}. This unique property has led to their potential practical applications such as in spintronic devices and quantum computation. However, identification of topological phases in general have been a challenging topic since they do not fall into the Landau paradigm of phase transitions due to the absence of a local order parameter. In recent years, machine learning techniques have offered a promising new approach to the search for such phases in non-interacting as well as interacting systems \cite{PhysRevLett.118.216401,PhysRevB.96.245119,PhysRevB.96.195145,PhysRevLett.120.066401,PhysRevB.98.085402,PhysRevB.97.115453,Shiina2020,Rodriguez-Nieva2019,Ming2019,https://doi.org/10.48550/arxiv.1901.03346,PhysRevLett.124.226401,Greplova_2020,PhysRevB.102.054512,PhysRevA.103.012419,PhysRevB.104.024506,PhysRevB.104.165108,10.21468/SciPostPhys.11.3.073,10.21468/SciPostPhys.14.1.005}. 

In order to characterize a three-dimensional TI one needs four $\mathbb{Z}_2$ topological indices $(\nu_0,\bm{\nu})$, comprising of one strong topological index $\nu_0$ and three weak topological indices $\bm{\nu}$. A strong topological insulator has a non-zero value of $\nu_0$, which ensures the existence of gapless surface states on all two-dimensional surfaces. In contrast, a weak topological insulator (WTI) has non-trivial values for some of the weak topological indices, but a zero strong topological index, resulting in gapless surface states only on some surfaces.

{The same idea can be generalized to two-dimensional (2D) WTI where the strong topological index $\nu_0$ vanishes while at least one of the two weak topological indices $(\nu_1,\nu_2)$ is non-zero. In this case, conducting edge states appear along certain boundaries of the system. An instance of a 2D WTI can be visualized in the following manner. A one-dimensional (1D) system invariant under the mirror symmetry class AI admits a non-zero $\mathbb{Z}$ topological index while for a 2D system under the same symmetry class, one has $\nu_0=0$ \cite{PhysRevB.88.075142}. If we now stack 1D AI mirror symmetric chains along the $y-$direction in such a manner that the resulting 2D system also belongs to the same symmetry class, we obtain a 2D WTI. The weak indices of the 2D system are given by the average of the topological indices of the constituent 1D chains and the conducting edge states appear along the two edges of the lattice parallel to the $y-$axis.}

In Ref.~\cite{10.21468/SciPostPhys.10.3.059}, such a WTI was observed when hard-core bosons (HCBs) were subjected to an on-site potential with alternating signs along the different $y$-layers of a periodic honeycomb lattice. With the help of Stochastic Series Expansion (SSE) quantum Monte Carlo (QMC) techniques supported by analytical calculations, the complete phase diagram of this model was obtained to consist of six different phases, including a WTI at density $1/4$ or $3/4$. 

In this work, we employ supervised machine learning techniques based on correlation functions, obtained from SSE QMC simulations, to study the phase diagram of the HCB model\footnote{{Although much work has been done regarding the application} {of machine learning methods to fermionic systems, there have been relatively fewer applications to hard-core boson systems. ML approaches have been applied to explicit HCB models using input data from QMC simulations \cite{https://doi.org/10.48550/arxiv.1707.00663,PhysRevB.99.121104} and to systems with 1D spinless fermions \cite{vanNieuwenburg2017} or spin-$1/2$ chains \cite{doi:10.1126/science.aag2302, PhysRevB.95.245134, Torlai2018, PhysRevB.98.104426, PhysRevLett.121.167204, PhysRevB.100.224202} that can be mapped to HCBs.}}. We train our ML model using the density-density correlation function and the spatial correlation function as input data, and compare their effectiveness in predicting the phase boundaries. Our results show that a straightforward ML architecture consisting of three convolutional neural network (CNN) layers is able to predict the transition points of the complicated phase diagram accurately for most of the cases. In one particular case, we also employed the learning-by-confusion unsupervised technique \cite{vanNieuwenburg2017} to analyze the performance of the supervised ML model. 

{Topological phases in the presence of interactions has been a topic of intense research in the past few decades \cite{PhysRevLett.100.156401, PhysRevB.81.085105, Pesin2010, PhysRevLett.106.100403, PhysRevLett.107.106402, PhysRevLett.107.166806, PhysRevB.87.085109, PhysRevB.87.085136}. The study of such systems are especially challenging since they require expensive numerical effort on large system sizes and the results often vary from mean-field expectations \cite{PhysRevB.88.245123, PhysRevB.88.075101, PhysRevB.89.035103, PhysRevB.92.085146, PhysRevB.92.085147}. In this regard, ML can offer useful insights \cite{10.21468/SciPostPhys.11.3.073, 10.21468/SciPostPhys.14.1.005, YU2023}. In Ref.~\cite{10.21468/SciPostPhys.10.3.059}, the WTI phase was argued to be robust against any finite amount of nearest-neighbor (NN) repulsion based on theoretical reasoning, since numerical computation of the topological invariant was expensive.}

{By training the ML model with correlation function data labeled by the topological invariant in the limit of zero NN repulsion, we predict the topological invariant in presence of strong NN repulsion, which is otherwise inaccessible numerically for a large enough system size. We find that our results support the theoretical arguments regarding the robustness of the WTI. Furthermore, by training the machine with the data obtained from the Hamiltonian without any explicit interaction, we predict the phase structure of the system in the presence of NN repulsive interactions. Interestingly, the ML model, trained in the regime without NN repulsion, is able to predict the phase boundaries of the interacting Hamiltonian successfully.} 

The paper is organized in the following manner. In Sec.~\ref{sec:hamil}, we discuss the HCB model Hamiltonian and the phase diagram. We explain the input data format that we use for machine learning in Sec.~\ref{sec:mldata}, which is followed by a discussion about the neural network architecture and its training process in Sec.~\ref{sec:mlarch}. Next, we discuss the results that we obtain from the ML model in Sec.~\ref{sec:results} and finally summarize our conclusions in Sec.~\ref{sec:conclusions}.

\section{Model Hamiltonian}\label{sec:hamil}
\begin{figure}[t]
\centering
\includegraphics[width=0.4\textwidth]{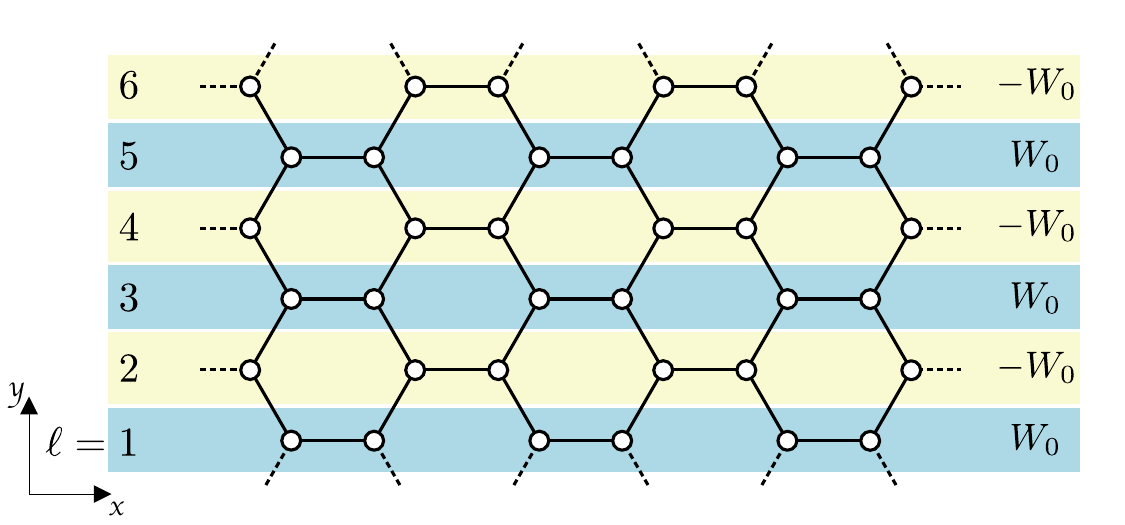}
\caption{Schematic diagram of the 2D periodic honeycomb lattice subjected to on-site potential $W_0$ with alternating signs along the different $y$-layers denoted by $l$. }
\label{fig:lattice}
\end{figure}

We study hard-core bosons on a two-dimensional (2D) periodic honeycomb lattice subjected to an on-site potential that has alternating signs along the $y$-layers of the lattice. The governing Hamiltonian can be expressed as,
\begin{align}\label{Hamiltonian}
    H=-t\sum_{\langle i,j\rangle}(\hat d_i^\dagger \hat d_j+\mathrm{h.c.})+\sum_i W_i \hat n_i-\sum_i \mu \hat n_i,
\end{align}
 where $\langle i,j\rangle$ denote nearest-neighbor (NN) pairs of sites, $\hat d_i^\dagger (\hat d_i)$ creates (annihilates) a HCB at site $i$ and $\hat n_i=\hat d_i^\dagger \hat d_i$ gives the number of HCBs at site $i$. In the above equation $t$ represents the hopping amplitude, $W_i$ denotes the on-site potential and $\mu$ stands for the chemical potential of the system. As depicted in Fig.~\ref{fig:lattice}, $W_i$ forms a periodic potential along the $y$-direction of the lattice with $W_i=W_0(-W_0)$ for layers labeled by odd (even) values of $l$.

\begin{figure}[t]
\centering
\includegraphics[width=0.45\textwidth]{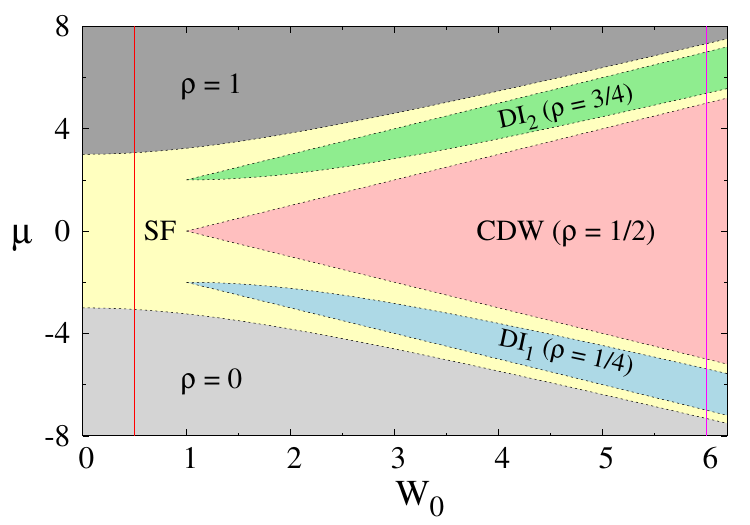}
	 \caption{Phase diagram of the Hamiltonian in Eq.~\ref{Hamiltonian} as a function of on-site potential $W_0$ and chemical potential $\mu$ {for a finite hopping, $t=1$}. The phase boundaries shown with dotted lines are obtained with analytical results in the thermodynamic limit \cite{10.21468/SciPostPhys.10.3.059}.}
\label{fig:phasediag}
\end{figure}

 In Ref.~\cite{10.21468/SciPostPhys.10.3.059}, this Hamiltonian was extensively studied using Stochastic Series Expansion (SSE) quantum Monte Carlo (QMC) simulations supported by analytical calculations and the complete phase diagram was constructed in terms of $t$, $W_0$ and $\mu$. {For finite hopping amplitude,} the system revealed the existence of six different phases : a charge density wave (CDW) at half-filling, two dimer insulators ($\mathrm{DI}_1$ and $\mathrm{DI}_2$) at densities $1/4$ and $3/4$ respectively, the empty phase at density $\rho=0$, the Mott insulator at density $\rho=1$ and a superfluid (SF) phase separating all these gapped phases from each other. {In Fig. \ref{fig:phasediag}, we show the phase diagram as a function of the on-site potential, $W_0$, and chemical potential, $\mu$, at fixed hopping, $t=1$. It should be noted that the CDW, $\mathrm{DI}_1$ and $\mathrm{DI}_2$ phases arise only for $|W_0|>t$.}
 
 It was demonstrated that depending on the sign of $W_0$ one of the dimer insulators becomes a weak topological insulator with a zero Chern number, but a non-trivial Berry phase. {In particular with a fixed choice of boundary, if for positive $W_0$ values, the dimer insulator at $1/4$-filling turns out to be the WTI, then for negative values of $W_0$ the dimer insulator at $3/4$-filling becomes the topological one.}
 The topological phase is protected by mirror-symmetry and belongs to the mirror-symmetry class AI {\cite{PhysRevB.88.075142}}. Furthermore it was shown that although the chiral symmetry of the Hamiltonian is broken in the presence of alternating on-site potential, an emergent partial chiral symmetry pins the edge states to the center of the WTI band.

 In this paper, using supervised ML techniques we have been able to reproduce the phase diagram of the Hamiltonian in Eq.~\ref{Hamiltonian}, with two types of correlation functions as input data generated using SSE QMC method. We have trained the ML model using the density-density and spatial correlation functions separately and compared their performance in predicting the phase boundaries. In order to capture the essential features of the phase diagram in Fig.~\ref{fig:phasediag}, we apply ML techniques along two different lines depicted by magenta and red solid lines. Along $W_0=0.5$ (red line), the system experiences two phase transitions between three phases as a function of $\mu$, whereas at $W_0=6.0$ (magenta line), the system goes through all six phases, including the WTI phase, through eight phase transitions. We compare the phase transition points predicted by our ML model with those obtained using exact analytical calculations \cite{10.21468/SciPostPhys.10.3.059}. 

Furthermore, in Ref.~\cite{10.21468/SciPostPhys.10.3.059} the effect of interactions between the HCBs on the phase diagram and especially on the WTI, was studied by introducing the following term in the Hamiltonian,
\begin{align}
    H_1=V_1\sum_{\langle i,j\rangle} \hat n_i \hat n_j,
\end{align}
where $V_1$ indicates the strength of the NN repulsion. 
Since it is numerically infeasible to calculate the topological invariant for a large enough interacting system, theoretical arguments were provided to show that the WTI is robust against any finite amount of NN repulsion. 

\begin{figure}[h]
	\includegraphics[width=0.8\linewidth]{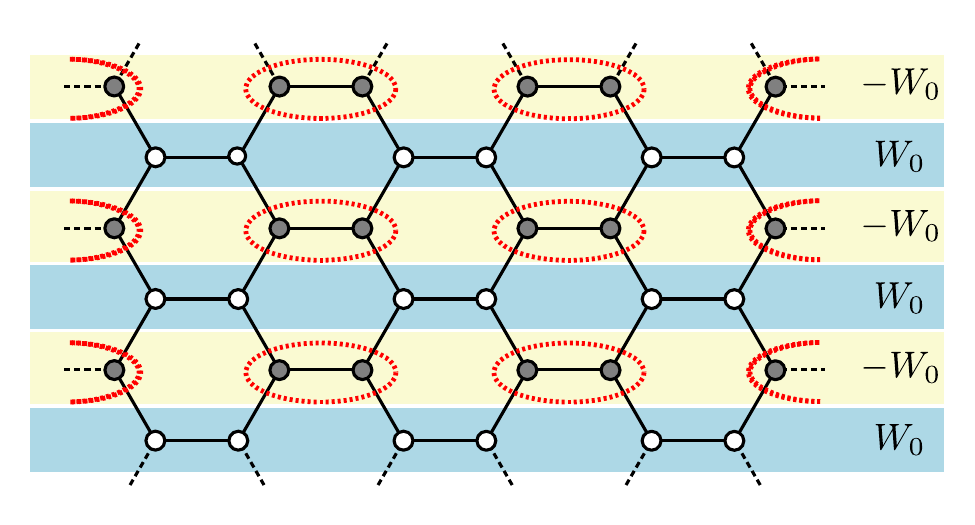}
	\caption{{Schematic lattice configuration of the $\mathrm{DI}_1$ dimer insulator phase at $\rho=1/4$ for positive $W_0$. The white and gray circles indicate lattice sites with particle density $0.0$ and $0.5$, respectively. A red dashed loop indicates a dimer with a single particle hopping back and forth between the two sites enclosed within.}}
	\label{fig:WTI_robust}
\end{figure}

{For completeness, we present here the reasoning for the robustness of the WTI which is based on the spatial structure of the dimer insulators. As depicted in the particle configuration of the WTI $\mathrm{DI}_1$ phase at density $\rho=1/4$ for positive $W_0$ in Fig.~\ref{fig:WTI_robust}, a particle hopping to and fro between the adjacent sites of a NN bond along the $x-$direction forms a dimer on alternate $y-$levels. Due to the absence of particles in the layers above and below, the particle in a dimer does not feel any repulsion regardless of the value of $V_1$ and the {back-and-forth} hopping process continues unhindered.} 

{For the case at $\rho=3/4$ for negative $W_0$ ($\mathrm{DI}_2$ phase), the dimers are still formed on the yellow $y-$levels shown in Fig.~\ref{fig:WTI_robust} but the blue layers are now completely filled. This is, in fact, the minimum energy configuration even when NN repulsion is turned on. Introducing $V_1$, the particle in each dimer now experiences the same repulsion $2V_1$, from the two occupied NN sites, irrespective of which end of the dimer it resides on. Therefore, instead of choosing a particular site, the particle in each dimer still prefers to hop back-and-forth thereby minimizing the energy of the system, while the particles in the layers in between remains frozen. In total, we see that the application of NN repulsion does not affect the dimer formation and the overall structure in the $\mathrm{DI}_1$ and $\mathrm{DI}_2$ phases, and thus the WTI phase stays robust.} 

{The effect of next-nearest-neighbor (NNN) repulsion on the system was further investigated in \cite{10.21468/SciPostPhys.10.3.059}. It was found that beyond some critical value of the NNN repulsion for a fixed value of $V_1$, the dimer formation is disrupted and the DI phases turn into a normal insulator. In this study, we only consider the NN repulsion term and we have tried to predict its effect on the phase boundaries as well as the topological invariant, with the ML model} trained in the non-interacting\footnote{{It should be kept in mind that the Hamiltonian in Eq.~\ref{Hamiltonian} is already interacting due to the hard-core nature of the particles. However, for brevity we call this Hamiltonian ``non-interacting'' and the one with explicit NN repulsion ``interacting''.}} regime.
 
 \begin{figure}[t]
\includegraphics[width=0.48\textwidth]{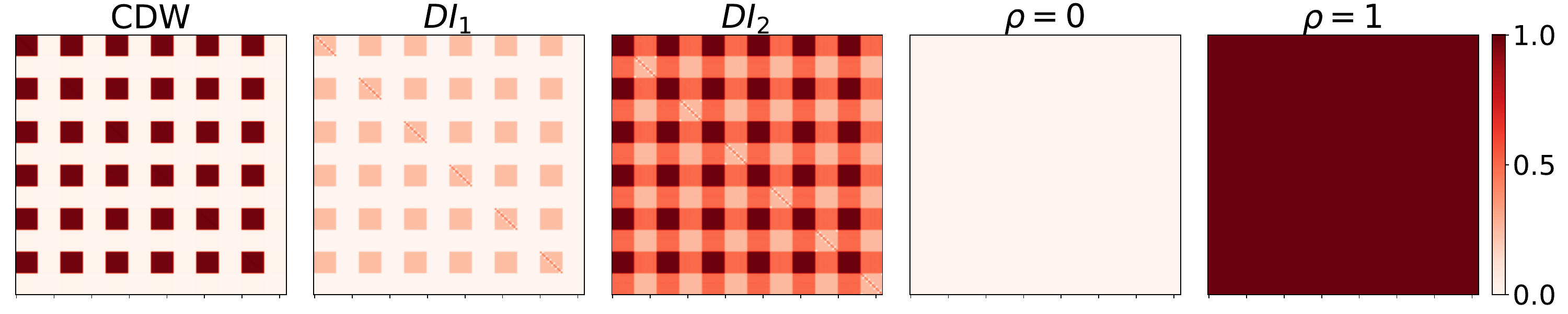}
\includegraphics[width=0.48\textwidth]{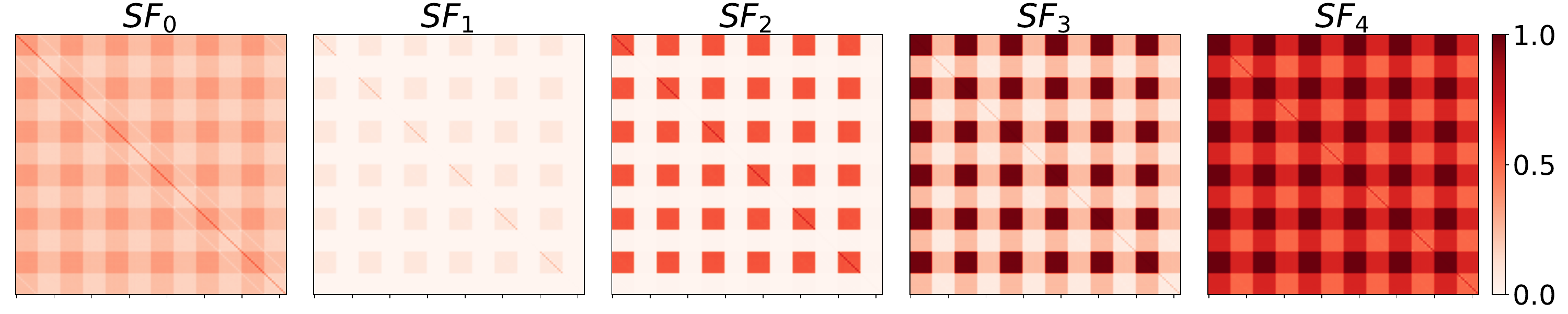}
\caption{Representative images of input data for DDC matrix obtained on a $12\times 12$ honeycomb lattice for six different phases - \textit{Top}: charge density wave (CDW), dimer insulator at density $1/4$ ($\mathrm{DI}_1$), dimer insulator at density $3/4$ ($\mathrm{DI}_2$), empty phase ($\rho=0$), Mott insulator ($\rho=1$) and, \textit{Bottom}: superfluid phase $\mathrm{SF}_0$ at $W_0=0.5$ close to half-filling, superfluid phases ($\mathrm{SF}_1$, $\mathrm{SF}_2$, $\mathrm{SF}_3$ and $\mathrm{SF}_4$) corresponding to densities $0<\rho<0.25$, $0.25<\rho<0.5$, $0.5<\rho<0.75$ and  $0.75<\rho<1$, respectively obtained, for a fixed $W_0=6.0$.}
\label{fig:data_image}
\end{figure}

\section{ML input data}\label{sec:mldata}
{Depending on the properties of the theory, different types of correlation functions can be chosen as input data for various ML applications. The phase diagram of our Hamiltonian consists of several insulating phases for which one observes plateaus in the particle density as a function of the chemical potential, $\mu$. Moreover, hopping plays an important role in various phases of the theory, for instance, the superfluid phase allows for hopping along a particular direction while the dimer insulator phases are characterized by the back-and-forth hopping of a single particle on a NN bond.}

As input of the ML model, we{, therefore,} consider two different types of correlation functions : density-density correlation and spatial correlation functions. The density-density correlation (DDC) function is defined as,
\begin{equation}
    D_{ij} = \langle\hat n_i \hat n_j\rangle,
\end{equation}
where $i$ and $j$ are any two sites of the honeycomb lattice and $\langle\cdots\rangle$ denotes the ensemble average. Using the SSE QMC technique \cite{PhysRevE.66.046701,10.1063/1.3518900} we calculate this quantity for each pair of sites on a $L_x \times L_y$ honeycomb lattice and store it as a $N\times N$ matrix, where $N=L_xL_y$ is the total number of lattice sites. These matrices are fed to the ML model as input. 

Next, the spatial correlation (SpC) function between two sites $i$ and $j$ is measured as
\begin{equation}
    S_{ij} = \langle\hat d_i^\dagger \hat d_j\rangle.
\end{equation}
This observable can be computed by introducing a modification in the usual loop construction in the directed loop update of SSE QMC, as proposed in Ref.~\cite{PhysRevE.64.066701}. {It may be noted that for hard-core bosons this observable is positive definite since HCB operators at different sites commute.}

\begin{figure}[t]
\includegraphics[width=0.48\textwidth]{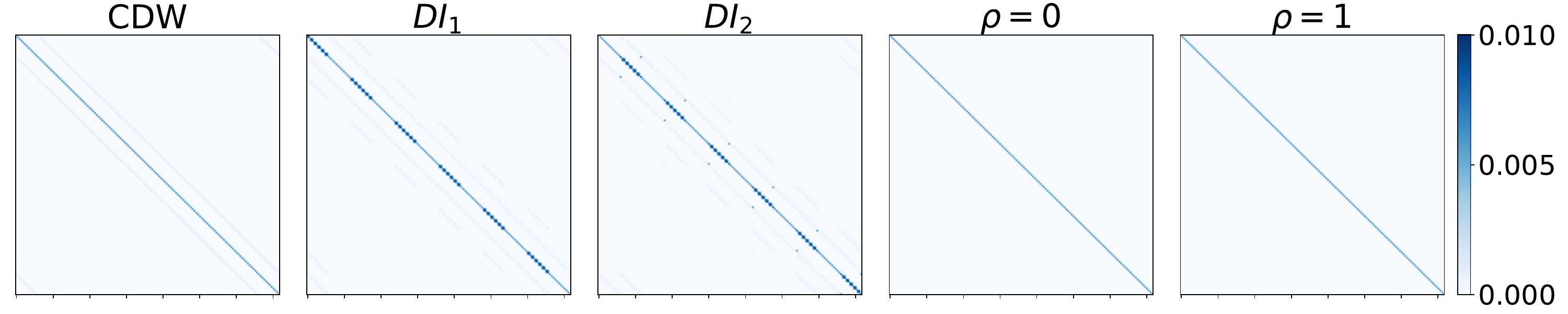}
\includegraphics[width=0.48\textwidth]{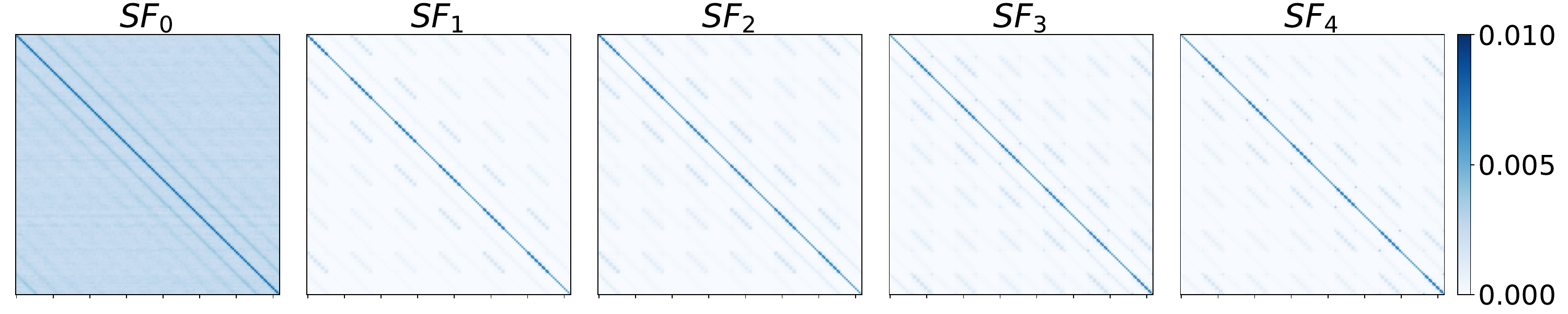}
\caption{Representative images of input data for SpC matrix obtained from similar regions of the phase diagram as in Fig.~\ref{fig:data_image} for a $12\times 12$ honeycomb lattice.}
\label{fig:data_image_Corr}
\end{figure}

Representative images from various phases corresponding to the DDC and SpC input data for system size $L\equiv L_x=L_y=12$ are depicted in Fig.~\ref{fig:data_image} and \ref{fig:data_image_Corr}. Although our main focus is on the system size $L=12$, we also investigate the dependence on system size by considering smaller lattices of $L=4$ and $8$. As seen in Fig.~\ref{fig:phasediag}, a superfluid region extending from densities $0<\rho<1$ divides into four branches for $W_0\geq 1$ corresponding to the following ranges of densities : $0<\rho<0.25$, $0.25<\rho<0.5$, $0.5<\rho<0.75$ and  $0.75<\rho<1$. For convenience, we designate the undivided SF region as $\mathrm{SF}_0$ and the divided regions as $\mathrm{SF}_1$, $\mathrm{SF}_2$, $\mathrm{SF}_3$ and $\mathrm{SF}_4$, respectively.

It is important to note that there is a major difference between the two correlation functions. In case of DDC function the input data depends on the average density of the system. Thus the images corresponding to six different phases are distinct. On the other hand, the SpC matrix, being invariant under particle-hole symmetry, is identical for the empty phase ($\rho=0$) and the Mott insulator phase ($\rho=1$), where all sites are empty and singly occupied, respectively.

Moreover, the spatial correlation function captures the emergent chiral symmetry around the center of the dimer insulator band for densities below (above) half-filling, which renders the images for superfluid regions $\mathrm{SF}_1$($\mathrm{SF}_3$) and $\mathrm{SF}_2$($\mathrm{SF}_4$) to be identical (see Fig.~\ref{fig:data_image_Corr}). However, the DDC images for the superfluid regions corresponding to different densities are distinct from one another.

\section{ML architecture and training}\label{sec:mlarch}
\begin{figure}[t]
\includegraphics[width=0.47\textwidth]{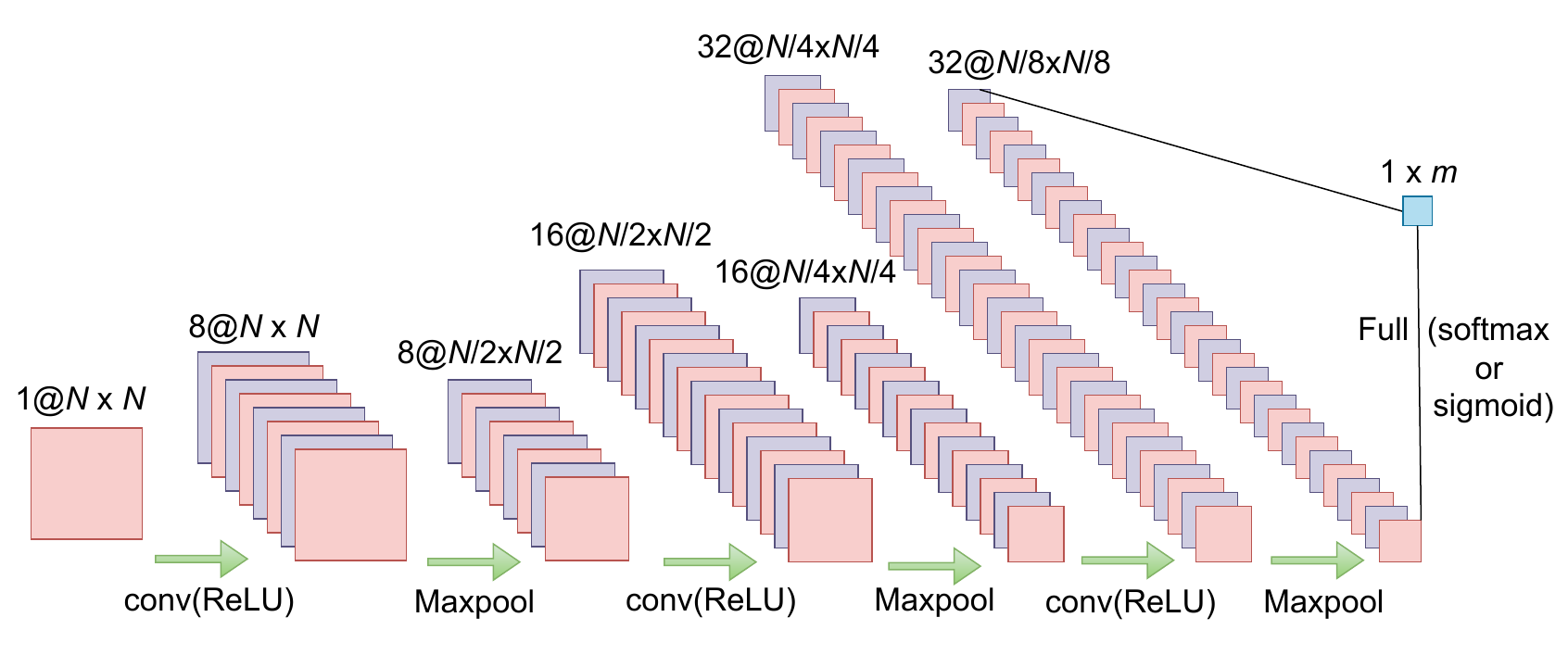}
\caption{Schematic diagram of the ML architecture used in our study. The number of channels and the dimension of the data in each channel is indicated above each layer.}
\label{fig:architecture}
\end{figure}

Taking advantage of the image-like representation of the input data, we use a convolutional neural network (CNN) model, which is best suited for image data classification tasks \cite{NIPS2012_c399862d,zeiler2013visualizing,sermanet2014overfeat,simonyan2015deep}. We use a relatively simple CNN architecture with only three CNN blocks followed by a fully connected output layer as depicted in Fig.~\ref{fig:architecture}. Each CNN block consists of a CNN layer with $k$ convolution filters followed by the non-linear ReLU activation function and a MaxPool layer of stride (2,2) which reduces each dimension of the 2D matrix by half. The first CNN layer is composed of $k=8$ convolution filters of size $3\times 3$ which acts on the single channel DDC/SpC input data, which are $144 \times 144$ matrices corresponding to a $12\times 12$ honeycomb lattice. For the interacting case, we provide the ML model with both DDC and SpC data at the same time in two separate channels. Motivated by the VGG16 model \cite{simonyan2015deep}, we keep doubling the number of filters as each dimension the 2D data reduces by half. The second and third CNN layers have $k=16$ and $32$ filters, respectively. The output from the last CNN block is flattened into an 1D array and fed to a fully-connected neural network layer consisting of $m$ neurons, depending on the prediction task.

For the prediction of the phase boundaries, the input data are labelled with one-hot encoded vectors of length $m$, where $m$ denotes the number of phases involved in the training process. Details about these labels for the DDC and SpC cases are discussed in Sec.~\ref{sec:phasepred}. On the other hand, for predicting the topological invariant, we use the value of the invariant as the label for input data. This means that the final layer of our neural network has only one neuron ($m=1$). When predicting phase boundaries, the output of the neurons in the last layer passes through the softmax activation function resulting in a final output which can be interpreted as a probability of belonging to a given phase. On the contrary, we use sigmoid activation function for the single neuron on the final layer for the prediction of topological invariant. 

\begin{figure}[t]
\centering
\includegraphics[width=0.35\textwidth]{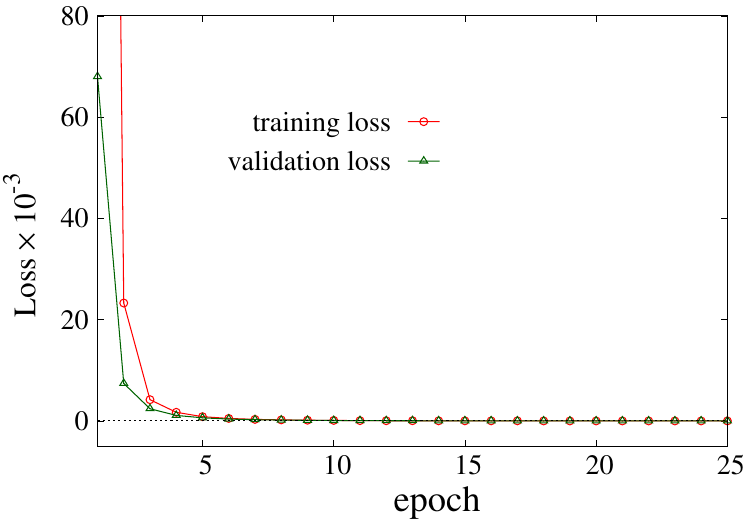}
\caption{Training and validation loss as a function of the number of epochs.}
\label{fig:loss}
\end{figure}

 To train our ML model for phase boundary detection, we utilize the categorical cross-entropy loss function minimized using the standard Adam optimization method \cite{kingma2017adam} with a learning rate of $0.0001$. For the prediction of the topological invariant, we use the binary cross-entropy loss function instead. For training, $2000$ data points were collected from deep within each phase of the phase diagram in Fig.~\ref{fig:phasediag} in a window of $\Delta\mu=0.04$ at fixed $W_0$. With a batchsize of 32, the training loss was found to decrease to below $10^{-6}$ in less than 20 epochs. The validation loss remained consistently below the training loss indicating the absence of overfitting. A typical example of the loss minimization with the number of training epochs is shown in Fig.~\ref{fig:loss}. The CNN model was implemented using the Keras framework \cite{chollet2015keras} in Python.

\section{Results and Discussion} \label{sec:results}

In this section, we describe the prediction results obtained from our ML model trained on the QMC data. The section is mainly divided into two parts. In the first part, we discuss the results of our ML models trained separately with DDC and SpC data obtained for the Hamiltonian in Eq.~\ref{Hamiltonian} without NN interactions. We contrast the performance of the two types of correlation functions in the prediction of the phase boundaries and compare them with the exact analytical results obtained in Ref.~\cite{10.21468/SciPostPhys.10.3.059}. Next, we apply the ML model trained with data from the non-interacting Hamiltonian to study the effect of interactions on the phase diagram, as well as on the topological invariant. In this case, the ML model is trained with the DDC and SpC data jointly.

\subsection{Phase diagram without NN interactions}\label{sec:phasepred}
 In order to reproduce the phase diagram of the Hamiltonian in Eq.~\ref{Hamiltonian}, the prediction data provided to the CNN model is taken along two distinct lines with fixed $W_0 = 6.0$ and $0.5$, highlighted by magenta and red lines, respectively, in Fig.~\ref{fig:phasediag}. For both cases, the prediction results have been obtained as the average of 200 independently trained models. We discuss below the results predicted by our CNN model along the above-mentioned two directions separately for training with both DDC and SpC data.

In the phase diagram shown in Fig.~\ref{fig:phasediag}, we can see that at $W_0=6.0$ the system undergoes eight phase transitions between six different phases - charge density wave (CDW), dimer insulator phases $\mathrm{DI}_1$ and $\mathrm{DI}_2$, superfluid phase (SF), the empty phase ($\rho=0$) and the Mott insulator phase ($\rho=1$), as the chemical potential of the system, $\mu$, is varied within $[-8,8]$. 
In order to train our ML model, labeled training datasets are prepared using SSE QMC methods, from deep within the aforementioned six phases away from any transition point. As can be seen from Fig. \ref{fig:data_image}, images of the DDC data corresponding to the four superfluid regions ($\mathrm{SF}_1$, $\mathrm{SF}_2$, $\mathrm{SF}_3$ and $\mathrm{SF}_4$) along the $W_0=6.0$ line appear to be different from each other. Therefore, we include data collected from the center of these four superfluid regions with the same label into the  DDC training dataset.

\begin{figure}[t]
\centering
\includegraphics[width=0.46\textwidth]{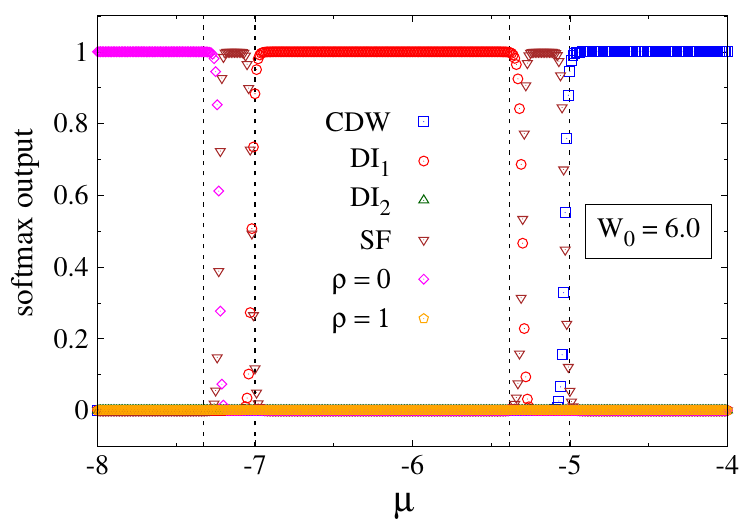}
\includegraphics[width=0.46\textwidth]{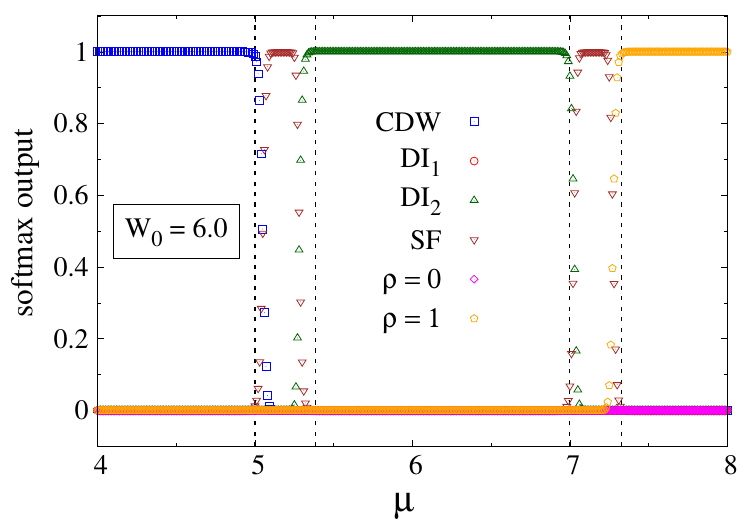}
\caption{Softmax neuron outputs from the ML model trained on DDC data as a function of chemical potential $\mu$ at fixed on-site potential $W_0=6.0$. The black dashed lines represent the transition points obtained from exact analytical calculations \cite{10.21468/SciPostPhys.10.3.059}.}
\label{fig:predict_N_Corr}
\end{figure}

In Fig.~\ref{fig:predict_N_Corr}, we display the output of the ML model's softmax layer when presented with previously unseen DDC data evenly-spaced along the range of $\mu=[-8,8]$.
The output layer of the ML model consists of six neurons corresponding to the six different phases and their output values indicate the probabilities of the system being in any one of them.
For the sake of clarity, the softmax neuron outputs of the ML model have been plotted separately in the regions of interest, $\mu=[-8,-4]$ and $\mu=[4,8]$.

From Fig.~\ref{fig:predict_N_Corr}, it can be clearly seen that for a fixed $W_0=6.0$ as the chemical potential $\mu$ of the system is varied, the ML model successfully realizes eight phase transitions, indicated by the intersections of the neuron outputs. For example in Fig.~\ref{fig:predict_N_Corr} top, the first transition takes place between the empty phase $(\rho=0)$ and the superfluid phase (SF). In this case the neuron output for the empty phase decreases from $1$ to $0$, while the neuron output corresponding to the SF phase increases from $0$ to $1$, giving rise to a crossing of these two curves. Throughout this transition the neuron outputs corresponding to other four phases remain vanishingly small. Similar description holds for the other transitions as well. In the intermediate region $\mu=[-4,4]$, the neuron output corresponding to the CDW phase remains close to unity while other outputs are close to zero. 

A phase transition point is characterized by the inability of the ML model to differentiate between two phases. As a result, at this point, the crossing neuron output curves have equal probabilities of $0.5$. After analyzing the intersections shown in Fig.~\ref{fig:predict_N_Corr}, we have compiled the transition points, averaged over 200 independently trained models, in Table \ref{table1}. Our ML model provides consistent results, as indicated by the standard deviation of the predicted transition points being less than $0.5\%$. We compare these with the transition points obtained from analytical results \cite{10.21468/SciPostPhys.10.3.059}. Remarkably, the ML model, trained on DDC data for a finite $12\times 12$ lattice, achieved impressive results by accurately predicting transition points with less than $2\%$ deviation from the exact infinite volume results. The success of the ML model in accurately predicting the boundaries of six different phases is particularly noteworthy given its relatively simple architecture. The model was able to extract information from correlation functions of all six phases simultaneously, which is an interesting observation.
\begin{figure}[t]
\centering
\includegraphics[width=0.46\textwidth]{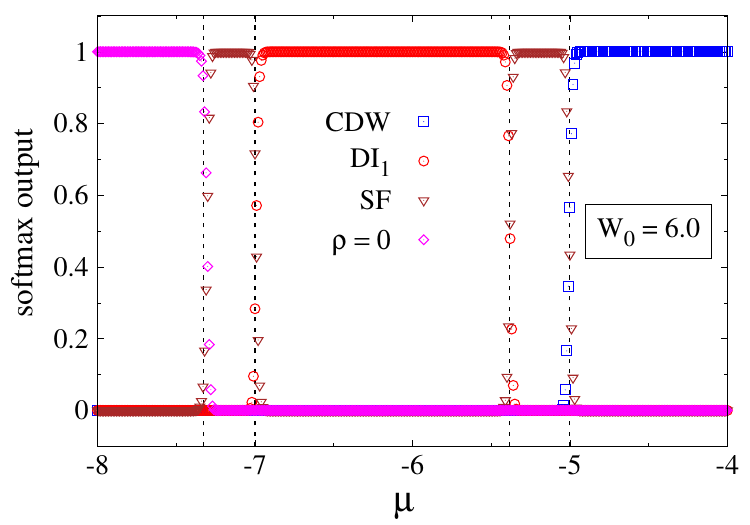}
\includegraphics[width=0.46\textwidth]{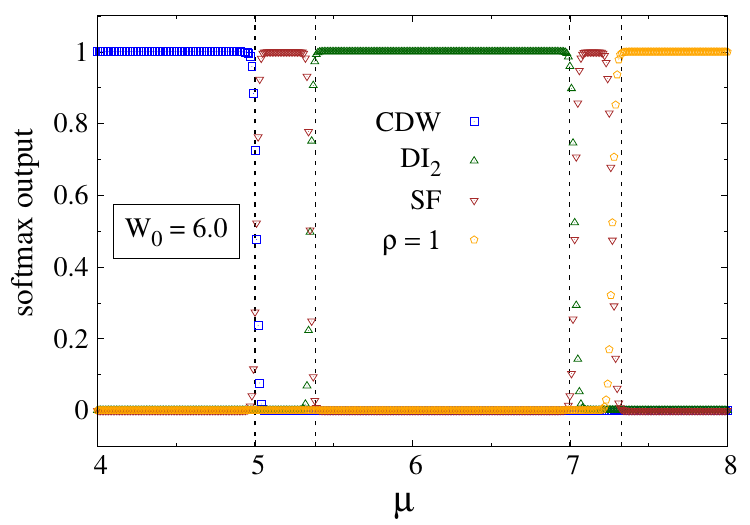}
\caption{Softmax neuron outputs of the ML model, trained on SpC data, plotted as a function of $\mu$ at fixed $W_0=6.0$. The positive and negative $\mu$ regions of the phase diagram have been obtained separately from two different trained models. The black dashed lines represent the exact transition points.}
\label{fig:predict_SP_Corr}
\end{figure}

Next, we trained the CNN model with spatial correlation (SpC) data obtained from QMC in the same region of the phase diagram as discussed above. Since the spatial correlation function is invariant under the particle-hole symmetry, the SpC data in the empty ($\rho=0$) and the Mott insulator ($\rho=1$) phases are exactly the same. This can also be seen in the representative images of Fig.~\ref{fig:data_image_Corr}. In order to avoid training with duplicate data having different phase labels, we divided the training and prediction process into two independent sets for positive and negative $\mu$ respectively at $W_0=6.0$. Each ML model is now trained with SpC data from four phases in either the positive or negative $\mu$ region of the phase diagram. As already mentioned in Sec.~\ref{sec:mldata}, there exists an emergent chiral symmetry in the system which renders the SpC data images for superfluid phases, $\mathrm{SF}_1$($\mathrm{SF}_3$) and $\mathrm{SF}_2$($\mathrm{SF}_4$) in the negative (positive) $\mu$ region to be identical. As a result, it is sufficient to train the ML model with SpC data from either superfluid phase, contrary to the training process with DDC data. Fig.~\ref{fig:predict_SP_Corr} shows the predictions of the ML models for positive and negative $\mu$, when supplied with unseen SpC data across the phase transitions. The predicted transition points again match impressively well with the analytical results, as shown in Table \ref{table1}. The standard deviation is less than $0.3\%$.

\begin{table}[t]
\centering
\begin{tabular}{|c|c|c|c|}
\hline
Transition between phases & Analytical & DDC & SPC \\ \hline
 ($\rho=0$) - SF & -7.325  & -7.227  &  -7.305 \\ \hline
 SF - $\mathrm{DI}_1$   &   -7.0  &  -7.020  & -6.992\\ \hline
    $\mathrm{DI}_1$ - SF  &   -5.385   & -5.302  & -5.380  \\ \hline
    SF - CDW     &  -5.0   & -5.032   & -5.003 \\ \hline
    CDW - SF     &   5.0  &  5.050  & 5.009 \\ \hline
    SF - $\mathrm{DI}_2$    &  5.385   &  5.283  & 5.350\\ \hline
    $\mathrm{DI}_2$ - SF    &   7.0   & 7.026  &  7.031 \\ \hline
    SF - ($\rho=1)$ & 7.325 & 7.274  & 7.269 \\ \hline
\end{tabular}
\caption{Comparison of transition points along $W_0=6$, obtained from the ML model, trained with DDC and SPC data, with exact analytical results.}
\label{table1}
\end{table}

\begin{figure}[b]
\centering
\includegraphics[width=0.4\textwidth]{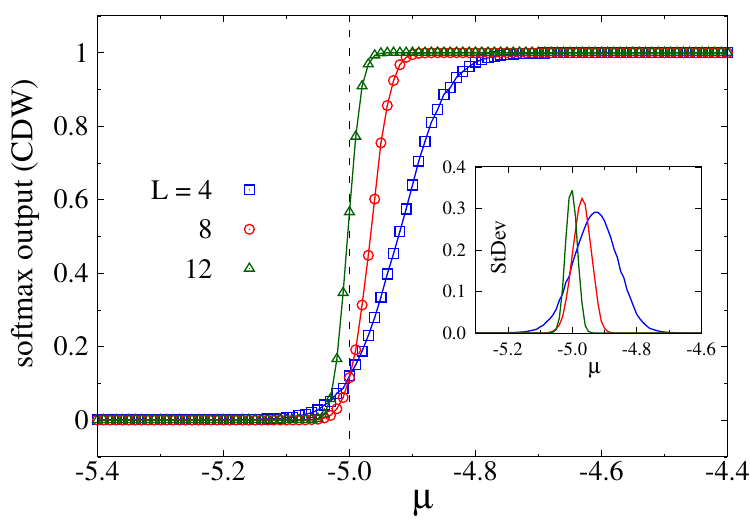}
	\caption{System size dependence of the softmax neuron output, corresponding to the CDW phase, as a function of $\mu$ at $W_0=6.0$. The training and prediction have been done with SpC data. The inset shows the standard deviation of the neuron output for the three system sizes. {The exact transition point is depicted by the black dashed line.}}
\label{fig:voldep}
\end{figure}

To understand the performance of our ML model on datasets of varying system sizes, we repeated the training and prediction process on DDC and SpC data obtained from QMC simulations on lattices with $L=4$ and $8$. In Fig.~\ref{fig:voldep}, we provide an example of the system size dependence of one of the softmax neuron outputs when trained with SpC data for three different system sizes. It clearly illustrates that the transition becomes steeper and approaches the exact transition point, indicated by the dashed vertical line, with increasing system size as one would expect. This is also evident from the standard deviation of the neuron output shown in the inset of the figure.  

\begin{figure}[t]
\centering
\includegraphics[width=0.46\textwidth]{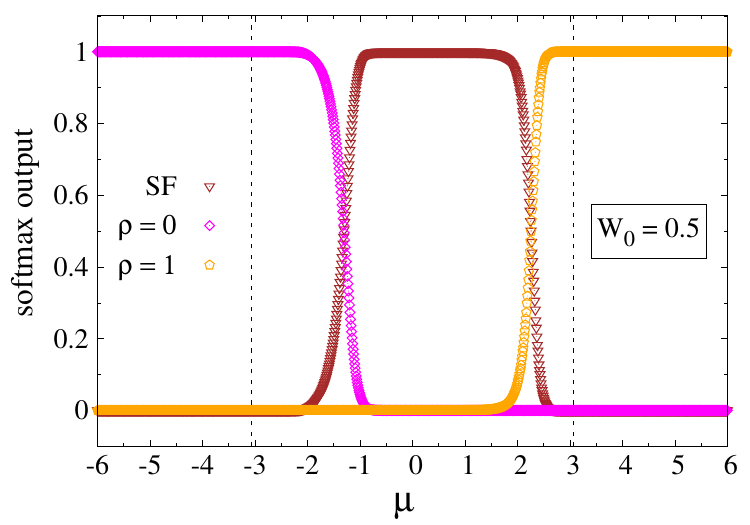}
\includegraphics[width=0.98\linewidth]{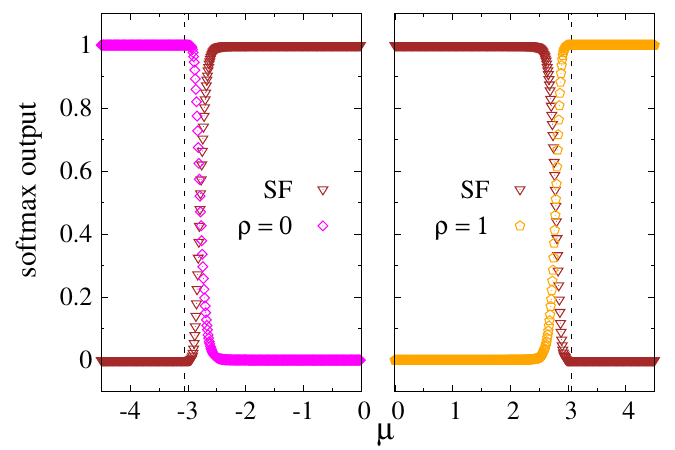}
	\caption{Prediction results at on-site potential $W_0=0.5$ as a function of $\mu$ obtained from, \textit{Top} : ML model trained with DDC data and, \textit{Bottom} : ML model trained with SpC data. The softmax output from the relevant positive and negative $\mu$ regions of the phase diagram have been obtained separately from two different trained models in the SpC case. {The black dashed lines display the exact transition points in the plots.}}
\label{fig:predict_N_Corr_W0p5}
\end{figure}

Next, we trained our ML model to learn information from the phases at a smaller value of on-site potential, $W_0=0.5$, which is highlighted by the red line in Fig.~\ref{fig:phasediag}. As $\mu$ is varied, the system goes through two phase transitions : from the empty phase ($\rho=0$) to the superfluid phase (SF) and therefrom to the Mott insulator phase ($\rho=1$). The two phase transitions occur symmetrically around $\mu=0$. Similar to the previous case of  $W_0=6.0$, the training data for both DDC and SpC have been taken from deep within the three phases in a very small window of $\Delta\mu=0.04$, and in particular, symmetrically around $\mu=0$ for the SF phase. In Fig.~\ref{fig:predict_N_Corr_W0p5} top, we show the output of the ML model, trained with DDC data, when provided with unseen data across the phase transitions. Although the ML model can distinguish the three phases, the predicted transition points have large deviations from the true values. Both the predicted transition points underestimate the superfluid phase and are also asymmetric around $\mu=0$. On the other hand, as seen in Fig.~\ref{fig:predict_N_Corr_W0p5} bottom, the ML model trained with SpC data performs significantly better with less than $10\%$ deviation from the analytical values of the transition points. It should be noted that the training with SpC data was done independently for the positive and negative $\mu$ regions, similar to the case of $W_0=6.0$. The analytical and predicted transition points, along $W_0=0.5$, for the DDC and SpC cases are listed in Table \ref{table2}. 

We tried to understand the reason behind the poor performance of the ML model trained with DDC data compared to the one with SpC data in the case of $W_0=0.5$. It is important to note that for on-site potential values $W_0<1$, the SF region persists over a wide range of densities, varying from $0$ to $1$. Since the density-density correlation function depends on the average density of the system, the DDC data for the SF phase shows a huge variation. Therefore, a supervised learning with training data collected from a small portion of the SF phase is insufficient to recognize a distinguishing pattern, which in turn results in the poor performance of the ML model. On the other hand, the spatial correlation function is related to the order parameter of a superfluid phase. Hence, the SpC data encode a unique pattern for the extended SF region which the machine can identify correctly even when trained in a small region. To check this hypothesis, we augmented the DDC training set with more data collected around $\mu=\pm1.5$ corresponding to average particle densities, around $1/4$ and $3/4$, and did the prediction again. This greatly improved the prediction of the transition points and thus, it shows that the poor performance was indeed due to lack of knowledge about the variation of DDC data in the SF phase. 

One should note that the ML model trained with DDC data had no difficulty in predicting the boundaries of the SF regions along $W_0=6.0$. At this value of the on-site potential, the range of densities over which the SF regions exist are significantly smaller compared to the $W_0=0.5$ case and hence, the DDC data does not have a large variation. As a result, the ML model performs significantly better. 

To further understand the features of the DDC data along $W_0=0.5$, we utilized the learning by confusion (LBC) method, an unsupervised ML technique introduced in Ref.~\cite{vanNieuwenburg2017} which tries to predict the phase transition point by deliberately confusing the ML model with fake data labels. For a dataset depending on some parameter $x\in[a,b]$ and characterizing a phase transition at some critical point $c$ in between, a partition point $c^\prime$ ($a<c^\prime<b$) is chosen and the dataset for $x\leq c^\prime$ is assigned the label $y(x\leq c^\prime)=0$ while $y(x>c^\prime)=1$. The ML model is trained with this fake-labelled dataset and the performance of the model $P(c^\prime)$ is assessed across the entire dataset. The function $P(c^\prime)$ exhibits an ubiquitous W-shaped curve with the central peak aligning with the true critical point at $c^\prime=c$.

The reason behind this W-shape can be understood in the following manner. On either side of the critical point, it is assumed that the ML model is able to recognize a set of unique features in the dataset. Setting $c^\prime=a$ ($c^\prime=b$) results in the ML model assigning label 1 (label 0) to both sets of features, thereby accurately predicting 100\% of the data. When $c^\prime=c$ and the training occurs with the true labels, the ML model correctly assigns appropriate labels leading to perfect performance. However for $a<c^\prime<c$ ($c<c^\prime<b$), the ML model encounters data with identical features but with different labels during the training, leading to confusion. In such instances, the ML model tends to learn the label of the majority data, impacting its performance according to \cite{vanNieuwenburg2017},
\begin{equation}\label{Eq:lbc_th}
P(c^\prime) = 1 - \frac{min(c-c^\prime,c^\prime-a)}{b-a}\,.
\end{equation}
This leads to the universal W-shape of $P(c^\prime)$ in the LBC scheme.

\begin{figure}[t]
\centering
\includegraphics[width=0.46\textwidth]{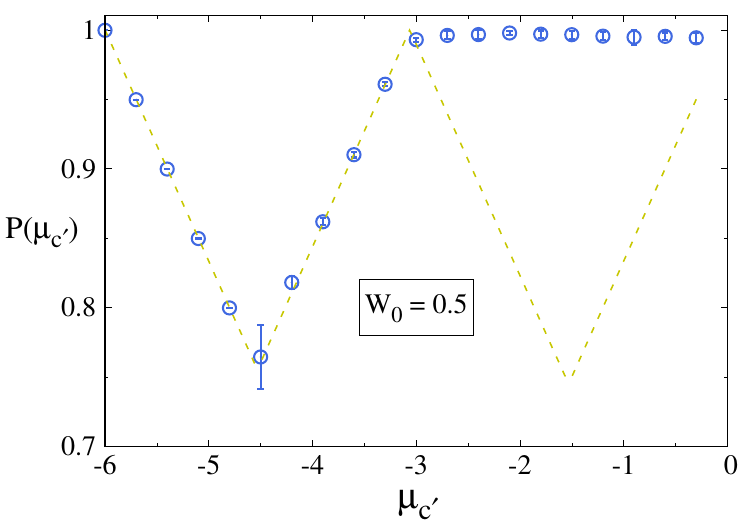}
\caption{Performance of the ML model as a function of partition point $\mu_{c^\prime}$
 in the LBC scheme. The ML model has been trained on the DDC data along $W_0=0.5$ in the range $\mu \in [-6,0]$. The error bars indicate the standard deviation over 10 independently trained models. Dashed line represents the expected behavior (Eq.~\ref{Eq:lbc_th})} \label{fig:lbc}.
\end{figure}    

We used the same CNN model architecture as depicted in Fig.~\ref{fig:architecture} and applied the LBC method to understand the DDC data along $W_0=0.5$ in the range $\mu \in [-6,0]$. For the performance metric $P(\mu_{c^\prime})$ as a function of the partition point $\mu_{c^\prime}$, we use the following quantity \cite{PhysRevX.12.031044},
\begin{equation}
P(\mu_{c^\prime}) = 1 - \frac{1}{N_D}\sum_{\mu}\left|\theta\left[\hat{y}_{c^\prime}(\mu)-0.5\right]-y_{c^\prime}(\mu)\right|\,,
\end{equation}
where $N_D$ is the total number of data points, $\theta$ is the Heaviside step function and $\hat{y}_{c^\prime}(\mu)$ and $y_{c^\prime}(\mu)$ denote the predicted labels and the assigned fake labels with respect to the partition point $\mu_{c^\prime}$, respectively. For the training, we have used a learning rate of 0.001 with a batchsize of 200 and evenly-spaced $N_D=3000$ data points. We have used a $l_2$ regularisation term with $l_2=0.01$ to minimize overfitting and prevent the ML model from focusing on smaller variations in the data \cite{vanNieuwenburg2017}. 

We show results for the performance function in Fig.~\ref{fig:lbc} and contrast it with the expected behavior given in Eq.~\ref{Eq:lbc_th} plotted as a dashed line. We find that the performance matches very well with the expected shape on the left side of the critical point $\mu_c=-3.062$ (\textit{cf}.~Tab.~\ref{table2}). Interestingly, on the right side of $\mu_c$, the ML model does not face any confusion with fake data labels corresponding to the partition points $\mu_c<\mu_c^\prime<0$ and the performance is always close to $1$. This behavior can be understood from the inability of the ML model to extract any distinguishing feature in the DDC data within the SF phase. Since the DDC data in the SF phase varies continuously, during the training process the ML model never encounters a situation where two or more data with same feature have different labels and hence faces no confusion.

\begin{table}[b]
\centering
\begin{tabular}{|c|c|c|c|}
\hline
Transition between phases & Analytical & DDC & SPC \\ \hline
 ($\rho=0$) - SF & -3.062  & -1.238  &  -2.786 \\ \hline
    SF - ($\rho=1)$ & 3.062 & 2.248  & 2.784 \\ \hline
\end{tabular}
\caption{Comparison of transition points at $W_0=0.5$, obtained from ML model, trained with DDC and SPC data, with exact analytical results.}
\label{table2}
\end{table}
\subsection{Effect of NN interactions}
We now proceed to investigate the HCB system with non-zero nearest-neighbor interactions using the ML model trained on the non-interacting dataset. Gaining experience about the pros and cons of the DDC and SpC input data, we decided to augment them together as a two-channel image, \textit{i.e.} a 3D tensor of shape $(N,N,2)$, for the training of the ML model. We use the same DDC and SpC data collected deep within the phases at $W_0=6.0$, which have been used for training in the non-interacting case discussed above in Sec.~\ref{sec:phasepred}.

\begin{figure}[t]
\centering
\includegraphics[width=0.46\textwidth]{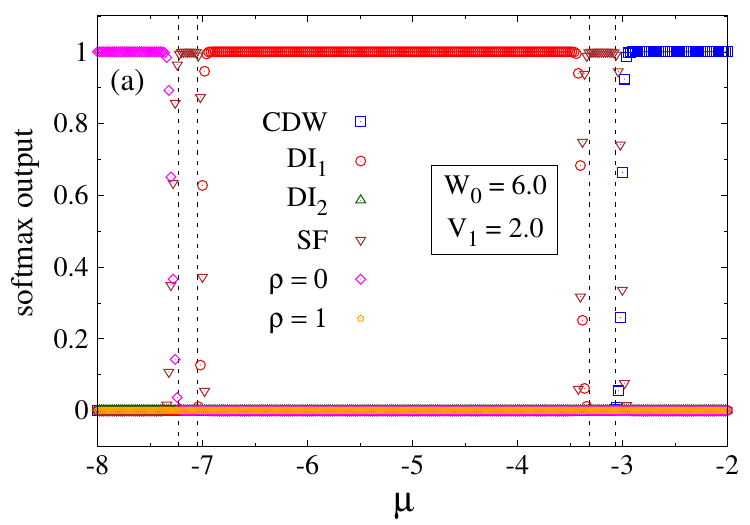}
\includegraphics[width=0.46\textwidth]{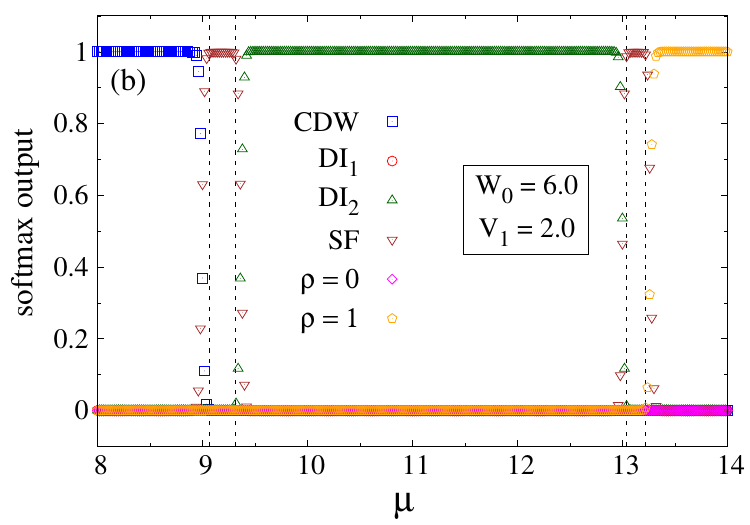}
\caption{Prediction of the ML model, trained on dual-channel DDC and SpC data obtained in the non-interacting case, plotted as a function of $\mu$ at on-site potential $W_0=6$ and NN repulsion $V_1=2$. The black dashed lines represent the transition points obtained from SSE QMC performed on a $12\times12$ honeycomb lattice.}
\label{fig:predict_V1_2}
\end{figure}

\begin{figure}[t]
\centering
\includegraphics[width=0.46\textwidth]{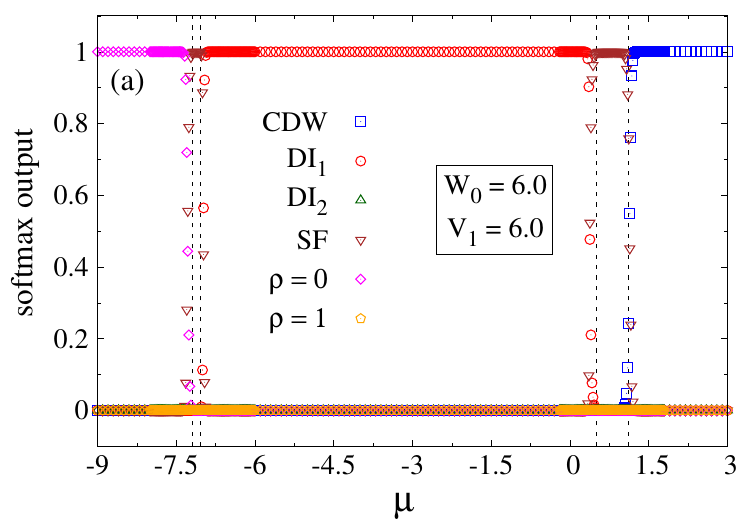}
\includegraphics[width=0.46\textwidth]{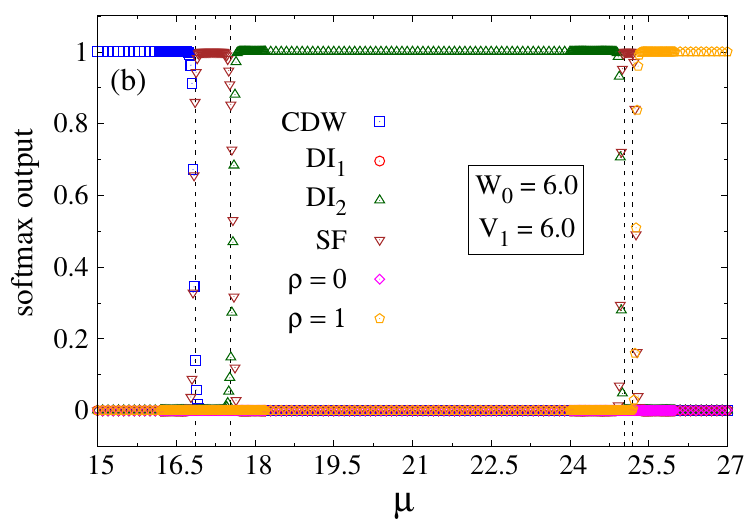}
\caption{Prediction on test data at NN repulsion $V_1=6$ obtained from the ML model, trained with non-interacting data, plotted as a function of $\mu$ at $W_0=6$.}
\label{fig:predict_V1_6}
\end{figure}

\subsubsection{Phase boundaries}
To predict the phase boundaries in the presence of NN interactions, we provide the ML model with unseen test data in the form of two-channel images of correlation functions, collected at repulsion strengths $V_1=2$ and $6$. These have been obtained using SSE QMC for varying chemical potential $\mu$ at $W_0=6$. In Ref.~\cite{10.21468/SciPostPhys.10.3.059}, it was observed that with increasing value of $V_1$, the band gaps corresponding to the gapped phases -- $\mathrm{DI}_1$ ($\rho=1/4$), CDW ($\rho=1/2$) and $\mathrm{DI}_2$ ($\rho=3/4$), get larger. In other words, the range of chemical potential over which these insulating phases exist increases in the presence of NN repulsion, and these phases become energetically more stable.

In Figs.~\ref{fig:predict_V1_2} and \ref{fig:predict_V1_6}, we show the predictions of our ML model at $V_1=2$ and $6$, respectively. For the sake of clarity, we have plotted the softmax neuron outputs of the ML model separately in the regions of interest, $\mu=[-8,-2]$ and $\mu=[8,14]$ for $V_1=2$ in Fig.~\ref{fig:predict_V1_2}, and $\mu=[-9,3]$ and $\mu=[15,27]$ for $V_1=6$ in Fig.~\ref{fig:predict_V1_6}, respectively. For $V_1=2$ ($V_1=6$), the CDW phase extends throughout the intermediate range of chemical potential $\mu=[-2,8]$ ($\mu=[3,15]$). As evident from these figures, the ML model correctly predicts the trend of increasing band gaps of the aforementioned three insulating phases with increasing $V_1$. The predicted transition points were determined by analyzing the intersection points of the softmax neuron output curves. We found that they were in good agreement with the transition points obtained from the QMC simulations for a $12\times12$ honeycomb lattice, as shown in Table \ref{tab:int_pt}. This demonstrates the ability of the ML model to learn the features from the non-interacting Hamiltonian and identify the phase structure in the interacting regime with good accuracy. 

\subsubsection{Topological invariant}
In Ref.~\cite{10.21468/SciPostPhys.10.3.059}, the topological nature of the different gapped phases at densities $\rho=1/4,1/2$ and $3/4$ were investigated by calculating the topological invariants, Chern number and Berry phase. For positive (negative) $W_0$ values, it was observed that the Chern number was zero in all three gapped phases, but a non-trivial Berry phase quantized at $\pi$ was obtained in the $\rho=1/4$ $\mathrm{DI}_1$ ($\rho=3/4$ $\mathrm{DI}_2$) phase. Therefore, the dimer insulator phase at $\rho=1/4$ ($\rho=3/4$) was identified to be a weak topological insulator for positive (negative) $W_0$ values.

\begin{table}[]
\begin{tabular}{|c|cc|cc|}
\hline
\multicolumn{1}{|c|}{\multirow{2}{*}{\begin{tabular}[c]{@{}c@{}}Transition \\ between phases\\ \end{tabular}}} & \multicolumn{2}{c|}{$V_1=2$} & \multicolumn{2}{c|}{$V_1=6$} \\ \cline{2-5} 
\multicolumn{1}{|c|}{} & \multicolumn{1}{c|}{QMC} & ML & \multicolumn{1}{c|}{QMC} & ML \\ \hline
 ($\rho=0$) - SF &  \multicolumn{1}{c|}{-7.224}  & -7.280  & \multicolumn{1}{c|}{-7.196} & -7.274 \\ \hline
 SF - $\mathrm{DI}_1$ & \multicolumn{1}{c|}{-7.044} & -7.004  & \multicolumn{1}{c|}{-7.042} & -6.982 \\ \hline
 $\mathrm{DI}_1$ - SF & \multicolumn{1}{c|}{-3.312} & -3.391 & \multicolumn{1}{c|}{0.498} & 0.384 \\ \hline
 SF - CDW & \multicolumn{1}{c|}{-3.068} & -3.008 & \multicolumn{1}{c|}{1.118} & 1.137 \\ \hline
 CDW - SF & \multicolumn{1}{c|}{9.070} & 8.994 & \multicolumn{1}{c|}{16.872} & 16.832 \\ \hline
 SF - $\mathrm{DI}_2$ & \multicolumn{1}{c|}{9.312} & 9.367  & \multicolumn{1}{c|}{17.526} & 17.578  \\ \hline
 $\mathrm{DI}_2$ - SF & \multicolumn{1}{c|}{13.042} & 13.002 & \multicolumn{1}{c|}{25.046} & 24.970 \\ \hline
 SF - ($\rho=1)$ & \multicolumn{1}{c|}{13.224} & 13.269 & \multicolumn{1}{c|}{25.196} & 25.260 \\ \hline
\end{tabular}
\caption{Transition points along $W_0=6$ with NN repulsion $V_1=2$ and $6$, obtained from SSE QMC simulations and from the ML model trained with combined DDC and SpC data.}
\label{tab:int_pt}
\end{table}

The computational effort of calculating the topological invariant in an interacting system scales exponentially with the system size. Therefore, based on theoretical arguments pertaining to the underlying structure of the insulator, it was shown that the WTI is robust against any finite amount of NN repulsion \cite{10.21468/SciPostPhys.10.3.059} {as discussed in Sec.~\ref{sec:hamil}}. In our study, we aim to overcome this numerical limitation by using ML techniques to predict the topological invariant of the system in presence of repulsive interactions.  

\begin{figure}[t]
\centering
\includegraphics[width=0.9\linewidth]{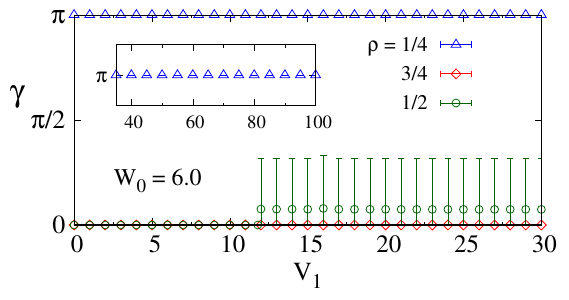}
\includegraphics[width=0.9\linewidth]{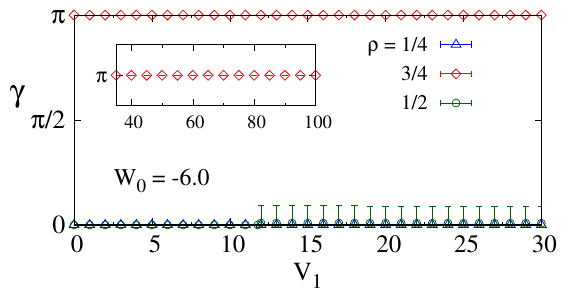}
	\caption{{Prediction results of Berry phase $\gamma$ as a function of NN repulsion $V_1$ for the CDW ($\rho=1/2$) and dimer insulator phases, $\mathrm {DI}_1$ ($\rho=1/4$) and $\mathrm{DI}_2$ ($\rho=3/4$), at fixed $W_0=6.0$ (\textit{Top}) and $-6.0$ (\textit{Bottom}). The error bars indicate the standard deviation of 200 independently trained ML models. It should be kept in mind that the ML prediction values are bounded between 0 and 1. The insets in both plots show the ML prediction of $\gamma$ at yet higher values of $V_1$ for the WTI phase, $\rho=1/4$ in \textit{top} and $\rho=3/4$ in \textit{bottom} respectively.}}
\label{fig:int_effect}
\end{figure}

We trained our model with dual channel correlation function data obtained in the non-interacting case at a fixed $W_0=6.0$. The input data from the various gapped phases are labelled with the value of the Berry phase $\gamma$, calculated in Ref.~\cite{10.21468/SciPostPhys.10.3.059}. This trained model was then used to predict the Berry phase on test data obtained using QMC in the interacting system. {In Fig.~\ref{fig:int_effect} \textit{top}, we show the prediction results for $\gamma$ as a function of the repulsion strength $V_1$. The machine clearly predicts that the Berry phase for the dimer insulator phase $\mathrm{DI_1}$ at density $\rho=1/4$ remains quantized at $\pi$, even in the presence of strong repulsive interactions.} This shows that the weak topological insulator phase remains unaffected by the presence of strong NN repulsive interactions and supports the argument for the robustness of the WTI phase.


{For the $\mathrm{DI}_2$ phase at $\rho=3/4$ in Fig.~\ref{fig:int_effect} \textit{top}, the topological invariant $\gamma$ predicted by the ML model remains close to zero for non-zero $V_1$, which demonstrates that the $\mathrm{DI}_2$ phase is unaffected by the NN repulsion as well. However, for the CDW phase at $\rho=1/2$, the machine predicts a vanishing value of $\gamma$ only for $V_1<12$. For $V_1>12$, the machine predicts a small non-zero value with a large standard deviation when averaged over 200 independently trained models. This result can be understood in the following manner.} 

\begin{figure}[t]
	\centering
	\includegraphics[width=0.9\linewidth]{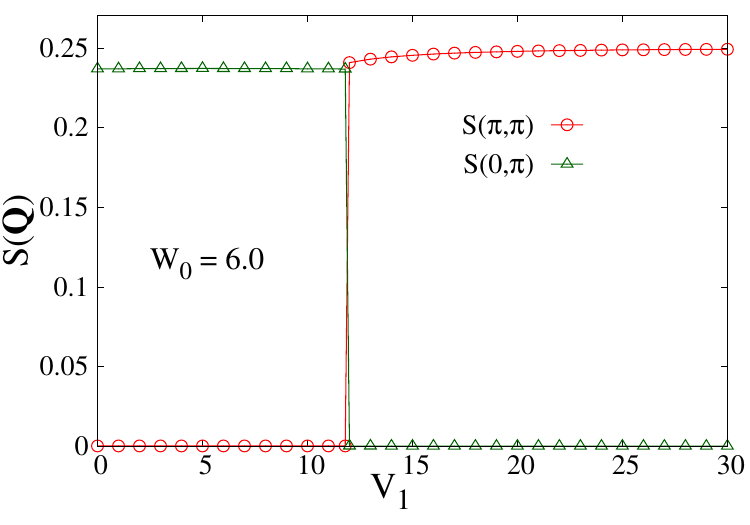}
	\includegraphics[width=0.9\linewidth]{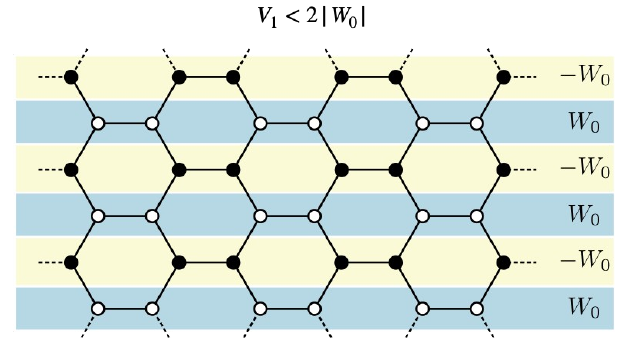}
	\includegraphics[width=0.9\linewidth]{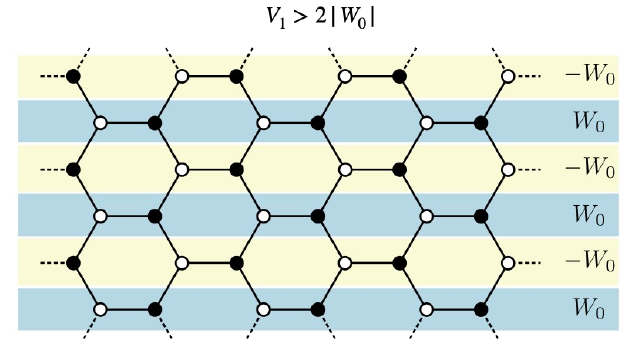}
	\caption{{\textit{Top}: Structure factors, $S(\pi,\pi)$ and $S(0,\pi)$, plotted as a function of the NN repulsion $V_1$ obtained from SSE QMC simulations on $12\times 12$ lattice. The corresponding CDW structures for $V_1<12$ (\textit{Middle}) and $V_1 \geq 12$ (\textit{Bottom}) are shown below. The white (black) circles indicate empty (occupied) lattice sites.}}
\label{fig:CDW_transition}
\end{figure}

{Although the increasing NN repulsion does not affect the $\mathrm{DI}_1$ and $\mathrm{DI}_2$ phases, the CDW structure at half-filling is, in fact, affected by it. As a function of $V_1$, a phase transition occurs from the original CDW of the non-interacting system (depicted in Fig.~\ref{fig:CDW_transition} \textit{middle}) to another CDW structure (depicted in Fig.~\ref{fig:CDW_transition} \textit{bottom}) in which either one of the two sublattices of the honeycomb lattice is occupied. The transition point can be determined by comparing the energies of the system in the two CDW states. In the original CDW phase at positive $W_0$, there are $N_s/2$ particles occupying the sites with on-site potential energy, $-W_0$ (depicted by the yellow layers in Fig.~\ref{fig:CDW_transition} \textit{middle}), and they form $N_s/4$ NN pairs experiencing $V_1 N_s/4$ repulsion leading to the total energy of the system being, $E_1=-W_0 N_s/2 + V_1 N_s/4$. On the other hand, in the second CDW phase, there is no repulsion felt between two HCBs as none of them reside on neighboring sites. Since half of the particles reside on sites with $-W_0$ on-site potential and the other half on sites with $W_0$ on-site potential, the total energy of the system, $E_2$, in this phase is zero. Therefore, the phase transition occurs at a critical value of the repulsion, $V_1=2W_0$, determined by setting $E_1=E_2$. A similar analysis can be done for the case of negative $W_0$ where the particles now occupy the blue layers of Fig.~\ref{fig:CDW_transition} \textit{middle} in the original CDW while the second CDW structure remains unchanged. Combining both the positive and negative $W_0$ cases, one obtains the critical repulsion to be $V_1=2|W_0|$.}


{We verified this transition with our SSE QMC simulations\footnote{{The emergence of this new CDW phase at large NN repulsion was not reported in Ref.~\cite{10.21468/SciPostPhys.10.3.059} since the system was studied up to $V_1=10$ at $W_0=6.0$.}} at $W_0=6.0$ as shown in Fig.~\ref{fig:CDW_transition} \textit{top}. We measured the structure factor per site, defined as,
\begin{equation}
	S(\bm{Q})=\frac{1}{N_s^2} \sum_{i,j} e^{i\bm{Q}\cdot(\bm{r}_i - \bm{r}_j)}\expval{\hat{n}_i \hat{n}_j} \, ,
\end{equation}
where $\bm{Q}$ denotes the wavevector and $\bm{r}_i$ are the coordinates at site $i$. The structure factor for wavevector $Q=(0,\pi)$ peaks in the original CDW phase while that for $Q=(\pi,\pi)$ peaks in the second CDW phase. In Fig.~\ref{fig:CDW_transition} \textit{top}, we find that $S(0,\pi)$ jumps down to zero at $V_1=2W_0=12$, with $S(\pi,\pi)$ shooting from 0 to its maximum value of 0.25 concomitantly.}

{Since the ML model has been trained on the input data obtained for the gapped phases of the non-interacting system, it runs into confusion when it encounters a new kind of data corresponding to the CDW structure at $V_1 \geq 2W_0$ for $W_0=6$. The machine outputs a wide range of values leading to a large standard deviation in comparison to the case at $\rho=1/4$ or $3/4$ (for all values of $V_1$) and $\rho=1/2$ for $V_1 < 12$, where the machine performs consistently. We repeated the same ML analysis for $W_0=-6.0$ as depicted in Fig.~\ref{fig:int_effect} \textit{bottom}. We find that the machine gets confused at the same point $V_1 \geq 12$ for $\rho=1/2$.
It may be noted that there is a difference in the mean values and standard deviations of the prediction of the machine for the positive and negative values of $W_0$ for $V_1 \geq 12$. This may be attributed to the fact that the machines were trained independently on two different sets of input data.}

{As mentioned in Sec.~\ref{sec:hamil}, the $\mathrm{DI}_2$ phase becomes topological for negative $W_0$ values in the non-interacting limit, while the $\mathrm{DI}_1$ phase is a trivial dimer insulator. As depicted in Fig.~\ref{fig:int_effect} \textit{bottom}, the machine now predicts the $\mathrm{DI}_2$ phase at $\rho=3/4$ to remain topological and the $\mathrm{DI}_1$ phase at $\rho=1/4$ to be trivial irrespective of the NN repulsion in agreement with the theoretical arguments.}

\section{Conclusions}\label{sec:conclusions}
To summarize, using a supervised machine learning approach, we have studied a system of hard-core bosons in a periodic honeycomb lattice subjected to on-site potential with alternating signs along different $y$-layers. This model was previously studied in detail using the SSE QMC method with additional analytical calculations \cite{10.21468/SciPostPhys.10.3.059}. With the application of ML techniques, in this paper we have reconstructed the rich phase diagram consisting of six different phases : a CDW at $1/2$-filling, two dimer insulators at densities $1/4$ and $3/4$, a superfluid phase, the empty phase at density $\rho=0$ and a Mott insulator at $\rho=1$. One of the dimer insulators, depending on the on-site potential pattern, is a weak topological insulator, with a zero Chern number but a non-trivial Berry phase.

{Based on the properties of the phases,} we use two different kinds of correlation functions, namely the density-density correlation (DDC) function and the spatial correlation (SpC) function, as our {ML} input data. The correlation functions were measured using the Stochastic Series Expansion QMC method. We train our ML model with the DDC and SpC data separately and compare their performance in learning the features encoded in the phase diagram. Along two distinct lines in the phase diagram, we determine the phase boundaries with the ML model and compare them with the ones obtained analytically \cite{10.21468/SciPostPhys.10.3.059}. In general, the phase transition points predicted by the ML model for a finite $12\times 12$ honeycomb lattice showed good agreement with the exact analytical results. However, the ML model was not able to correctly learn the features when trained with DDC data from a narrow region within the superfluid phase along one of the lines in the phase diagram. We performed an unsupervised analysis of the DDC data in this region using the learning by confusion approach which indicated that the DDC data shows large variation in the superfluid phase.

Next, we proceeded to examine the impact of {explicit} interactions between the HCBs on the phase diagram. For this purpose, we employed the ML model, which was trained with correlation functions obtained from {the Hamiltonian without any explicit interactions}, to investigate the phase structure of the system in the presence of nearest-neighbor repulsive interactions. In this scenario, we trained the ML model with DDC and SpC data as two-channel images. Notably, the ML model exhibited the ability to accurately predict the phase boundaries of the interacting Hamiltonian, despite being trained in the non-interacting regime. The capability of the ML model to learn features from the non-interacting system and predict them in the interacting case makes it particularly valuable for cases where numerical simulations in the interacting theory are challenging.

Furthermore, in Ref.~\cite{10.21468/SciPostPhys.10.3.059}, it was demonstrated that the weak topological insulator phase is highly stable against any finite amount of NN repulsion, but the topological invariant could not be calculated in the presence of interactions due to numerical limitations. As a result, the stability of the WTI phase was instead supported by theoretical arguments. This study aims to bridge this gap by utilizing ML techniques to predict the topological invariant as a function of repulsive interactions. By training the model on the non-interacting case, the machine predicts that the topological invariant remains unchanged even {in the presence of very large NN repulsion.} 

{However, the CDW phase at half-filling undergoes a phase transition to a different CDW phase at some critical value of the NN repulsion for a given value of the on-site potential. In the presence of NN repulsion, the ML model trained in the non-interacting regime correctly identifies the original CDW phase, but runs into confusion once it encounters the new CDW phase and outputs a wide range of values.}

\begin{acknowledgments}
This work was supported by the National Science and Technology Council, the Ministry of Education (Higher Education Sprout Project NTU-112L104022), and the National Center for Theoretical Sciences of Taiwan. It has also been supported in part by the Taiwanese  NSTC project 112-2639-M-002-006-ASP; Center for Theory and Computation, National Tsing Hua University-113H0001L9, Hsinchu 300 Taiwan. AG would like to thank M.~C.~Chung for introducing her to the concept of using correlation functions as ML input. The authors would also like to thank Julian Arnold for helpful comments. 
\end{acknowledgments}


\begin{thebibliography}{71}%
\makeatletter
\providecommand \@ifxundefined [1]{%
 \@ifx{#1\undefined}
}%
\providecommand \@ifnum [1]{%
 \ifnum #1\expandafter \@firstoftwo
 \else \expandafter \@secondoftwo
 \fi
}%
\providecommand \@ifx [1]{%
 \ifx #1\expandafter \@firstoftwo
 \else \expandafter \@secondoftwo
 \fi
}%
\providecommand \natexlab [1]{#1}%
\providecommand \enquote  [1]{``#1''}%
\providecommand \bibnamefont  [1]{#1}%
\providecommand \bibfnamefont [1]{#1}%
\providecommand \citenamefont [1]{#1}%
\providecommand \href@noop [0]{\@secondoftwo}%
\providecommand \href [0]{\begingroup \@sanitize@url \@href}%
\providecommand \@href[1]{\@@startlink{#1}\@@href}%
\providecommand \@@href[1]{\endgroup#1\@@endlink}%
\providecommand \@sanitize@url [0]{\catcode `\\12\catcode `\$12\catcode `\&12\catcode `\#12\catcode `\^12\catcode `\_12\catcode `\%12\relax}%
\providecommand \@@startlink[1]{}%
\providecommand \@@endlink[0]{}%
\providecommand \url  [0]{\begingroup\@sanitize@url \@url }%
\providecommand \@url [1]{\endgroup\@href {#1}{\urlprefix }}%
\providecommand \urlprefix  [0]{URL }%
\providecommand \Eprint [0]{\href }%
\providecommand \doibase [0]{https://doi.org/}%
\providecommand \selectlanguage [0]{\@gobble}%
\providecommand \bibinfo  [0]{\@secondoftwo}%
\providecommand \bibfield  [0]{\@secondoftwo}%
\providecommand \translation [1]{[#1]}%
\providecommand \BibitemOpen [0]{}%
\providecommand \bibitemStop [0]{}%
\providecommand \bibitemNoStop [0]{.\EOS\space}%
\providecommand \EOS [0]{\spacefactor3000\relax}%
\providecommand \BibitemShut  [1]{\csname bibitem#1\endcsname}%
\let\auto@bib@innerbib\@empty
\bibitem [{\citenamefont {Dawid}\ \emph {et~al.}(2022)\citenamefont {Dawid} \emph {et~al.}}]{Dawid:2022fga}%
  \BibitemOpen
  \bibfield  {author} {\bibinfo {author} {\bibfnamefont {A.}~\bibnamefont {Dawid}} \emph {et~al.},\ }\bibfield  {title} {\bibinfo {title} {{Modern applications of machine learning in quantum sciences}}\ }(\bibinfo {year} {2022})\ \Eprint {https://arxiv.org/abs/2204.04198} {arXiv:2204.04198 [quant-ph]} \BibitemShut {NoStop}%
\bibitem [{\citenamefont {Bedolla}\ \emph {et~al.}(2020)\citenamefont {Bedolla}, \citenamefont {Padierna},\ and\ \citenamefont {Castañeda-Priego}}]{Bedolla_2021}%
  \BibitemOpen
  \bibfield  {author} {\bibinfo {author} {\bibfnamefont {E.}~\bibnamefont {Bedolla}}, \bibinfo {author} {\bibfnamefont {L.~C.}\ \bibnamefont {Padierna}},\ and\ \bibinfo {author} {\bibfnamefont {R.}~\bibnamefont {Castañeda-Priego}},\ }\bibfield  {title} {\bibinfo {title} {Machine learning for condensed matter physics},\ }\href {https://doi.org/10.1088/1361-648X/abb895} {\bibfield  {journal} {\bibinfo  {journal} {Journal of Physics: Condensed Matter}\ }\textbf {\bibinfo {volume} {33}},\ \bibinfo {pages} {053001} (\bibinfo {year} {2020})}\BibitemShut {NoStop}%
\bibitem [{\citenamefont {Carleo}\ \emph {et~al.}(2019)\citenamefont {Carleo}, \citenamefont {Cirac}, \citenamefont {Cranmer}, \citenamefont {Daudet}, \citenamefont {Schuld}, \citenamefont {Tishby}, \citenamefont {Vogt-Maranto},\ and\ \citenamefont {Zdeborov\'a}}]{RevModPhys.91.045002}%
  \BibitemOpen
  \bibfield  {author} {\bibinfo {author} {\bibfnamefont {G.}~\bibnamefont {Carleo}}, \bibinfo {author} {\bibfnamefont {I.}~\bibnamefont {Cirac}}, \bibinfo {author} {\bibfnamefont {K.}~\bibnamefont {Cranmer}}, \bibinfo {author} {\bibfnamefont {L.}~\bibnamefont {Daudet}}, \bibinfo {author} {\bibfnamefont {M.}~\bibnamefont {Schuld}}, \bibinfo {author} {\bibfnamefont {N.}~\bibnamefont {Tishby}}, \bibinfo {author} {\bibfnamefont {L.}~\bibnamefont {Vogt-Maranto}},\ and\ \bibinfo {author} {\bibfnamefont {L.}~\bibnamefont {Zdeborov\'a}},\ }\bibfield  {title} {\bibinfo {title} {Machine learning and the physical sciences},\ }\href {https://doi.org/10.1103/RevModPhys.91.045002} {\bibfield  {journal} {\bibinfo  {journal} {Rev. Mod. Phys.}\ }\textbf {\bibinfo {volume} {91}},\ \bibinfo {pages} {045002} (\bibinfo {year} {2019})}\BibitemShut {NoStop}%
\bibitem [{\citenamefont {Carrasquilla}(2020)}]{doi:10.1080/23746149.2020.1797528}%
  \BibitemOpen
  \bibfield  {author} {\bibinfo {author} {\bibfnamefont {J.}~\bibnamefont {Carrasquilla}},\ }\bibfield  {title} {\bibinfo {title} {Machine learning for quantum matter},\ }\href {https://doi.org/10.1080/23746149.2020.1797528} {\bibfield  {journal} {\bibinfo  {journal} {Advances in Physics: X}\ }\textbf {\bibinfo {volume} {5}},\ \bibinfo {pages} {1797528} (\bibinfo {year} {2020})},\ \Eprint {https://arxiv.org/abs/https://doi.org/10.1080/23746149.2020.1797528} {https://doi.org/10.1080/23746149.2020.1797528} \BibitemShut {NoStop}%
\bibitem [{\citenamefont {Wang}(2016)}]{PhysRevB.94.195105}%
  \BibitemOpen
  \bibfield  {author} {\bibinfo {author} {\bibfnamefont {L.}~\bibnamefont {Wang}},\ }\bibfield  {title} {\bibinfo {title} {Discovering phase transitions with unsupervised learning},\ }\href {https://doi.org/10.1103/PhysRevB.94.195105} {\bibfield  {journal} {\bibinfo  {journal} {Phys. Rev. B}\ }\textbf {\bibinfo {volume} {94}},\ \bibinfo {pages} {195105} (\bibinfo {year} {2016})}\BibitemShut {NoStop}%
\bibitem [{\citenamefont {Ohtsuki}\ and\ \citenamefont {Ohtsuki}(2016)}]{doi:10.7566/JPSJ.85.123706}%
  \BibitemOpen
  \bibfield  {author} {\bibinfo {author} {\bibfnamefont {T.}~\bibnamefont {Ohtsuki}}\ and\ \bibinfo {author} {\bibfnamefont {T.}~\bibnamefont {Ohtsuki}},\ }\bibfield  {title} {\bibinfo {title} {Deep learning the quantum phase transitions in random two-dimensional electron systems},\ }\href {https://doi.org/10.7566/JPSJ.85.123706} {\bibfield  {journal} {\bibinfo  {journal} {Journal of the Physical Society of Japan}\ }\textbf {\bibinfo {volume} {85}},\ \bibinfo {pages} {123706} (\bibinfo {year} {2016})}\BibitemShut {NoStop}%
\bibitem [{\citenamefont {Carrasquilla}\ and\ \citenamefont {Melko}(2017)}]{carrasquilla_machine_2017}%
  \BibitemOpen
  \bibfield  {author} {\bibinfo {author} {\bibfnamefont {J.}~\bibnamefont {Carrasquilla}}\ and\ \bibinfo {author} {\bibfnamefont {R.~G.}\ \bibnamefont {Melko}},\ }\bibfield  {title} {\bibinfo {title} {Machine learning phases of matter},\ }\href {https://doi.org/10.1038/nphys4035} {\bibfield  {journal} {\bibinfo  {journal} {Nature Physics}\ }\textbf {\bibinfo {volume} {13}},\ \bibinfo {pages} {431} (\bibinfo {year} {2017})},\ \bibinfo {note} {number: 5 Publisher: Nature Publishing Group}\BibitemShut {NoStop}%
\bibitem [{\citenamefont {van Nieuwenburg}\ \emph {et~al.}(2017)\citenamefont {van Nieuwenburg}, \citenamefont {Liu},\ and\ \citenamefont {Huber}}]{vanNieuwenburg2017}%
  \BibitemOpen
  \bibfield  {author} {\bibinfo {author} {\bibfnamefont {E.}~\bibnamefont {van Nieuwenburg}}, \bibinfo {author} {\bibfnamefont {Y.-H.}\ \bibnamefont {Liu}},\ and\ \bibinfo {author} {\bibfnamefont {S.}~\bibnamefont {Huber}},\ }\bibfield  {title} {\bibinfo {title} {Learning phase transitions by confusion},\ }\href {https://doi.org/10.1038/nphys4037} {\bibfield  {journal} {\bibinfo  {journal} {Nature Physics}\ }\textbf {\bibinfo {volume} {13}},\ \bibinfo {pages} {435} (\bibinfo {year} {2017})}\BibitemShut {NoStop}%
\bibitem [{\citenamefont {Tanaka}\ and\ \citenamefont {Tomiya}(2017)}]{doi:10.7566/JPSJ.86.063001}%
  \BibitemOpen
  \bibfield  {author} {\bibinfo {author} {\bibfnamefont {A.}~\bibnamefont {Tanaka}}\ and\ \bibinfo {author} {\bibfnamefont {A.}~\bibnamefont {Tomiya}},\ }\bibfield  {title} {\bibinfo {title} {Detection of phase transition via convolutional neural networks},\ }\href {https://doi.org/10.7566/JPSJ.86.063001} {\bibfield  {journal} {\bibinfo  {journal} {Journal of the Physical Society of Japan}\ }\textbf {\bibinfo {volume} {86}},\ \bibinfo {pages} {063001} (\bibinfo {year} {2017})}\BibitemShut {NoStop}%
\bibitem [{\citenamefont {Wetzel}\ and\ \citenamefont {Scherzer}(2017)}]{PhysRevB.96.184410}%
  \BibitemOpen
  \bibfield  {author} {\bibinfo {author} {\bibfnamefont {S.~J.}\ \bibnamefont {Wetzel}}\ and\ \bibinfo {author} {\bibfnamefont {M.}~\bibnamefont {Scherzer}},\ }\bibfield  {title} {\bibinfo {title} {Machine learning of explicit order parameters: From the ising model to su(2) lattice gauge theory},\ }\href {https://doi.org/10.1103/PhysRevB.96.184410} {\bibfield  {journal} {\bibinfo  {journal} {Phys. Rev. B}\ }\textbf {\bibinfo {volume} {96}},\ \bibinfo {pages} {184410} (\bibinfo {year} {2017})}\BibitemShut {NoStop}%
\bibitem [{\citenamefont {Hu}\ \emph {et~al.}(2017)\citenamefont {Hu}, \citenamefont {Singh},\ and\ \citenamefont {Scalettar}}]{PhysRevE.95.062122}%
  \BibitemOpen
  \bibfield  {author} {\bibinfo {author} {\bibfnamefont {W.}~\bibnamefont {Hu}}, \bibinfo {author} {\bibfnamefont {R.~R.~P.}\ \bibnamefont {Singh}},\ and\ \bibinfo {author} {\bibfnamefont {R.~T.}\ \bibnamefont {Scalettar}},\ }\bibfield  {title} {\bibinfo {title} {Discovering phases, phase transitions, and crossovers through unsupervised machine learning: A critical examination},\ }\href {https://doi.org/10.1103/PhysRevE.95.062122} {\bibfield  {journal} {\bibinfo  {journal} {Phys. Rev. E}\ }\textbf {\bibinfo {volume} {95}},\ \bibinfo {pages} {062122} (\bibinfo {year} {2017})}\BibitemShut {NoStop}%
\bibitem [{\citenamefont {Broecker}\ \emph {et~al.}(2017)\citenamefont {Broecker}, \citenamefont {Assaad},\ and\ \citenamefont {Trebst}}]{https://doi.org/10.48550/arxiv.1707.00663}%
  \BibitemOpen
  \bibfield  {author} {\bibinfo {author} {\bibfnamefont {P.}~\bibnamefont {Broecker}}, \bibinfo {author} {\bibfnamefont {F.~F.}\ \bibnamefont {Assaad}},\ and\ \bibinfo {author} {\bibfnamefont {S.}~\bibnamefont {Trebst}},\ }\href {https://doi.org/10.48550/ARXIV.1707.00663} {\bibinfo {title} {Quantum phase recognition via unsupervised machine learning}} (\bibinfo {year} {2017})\BibitemShut {NoStop}%
\bibitem [{\citenamefont {Ch'ng}\ \emph {et~al.}(2018)\citenamefont {Ch'ng}, \citenamefont {Vazquez},\ and\ \citenamefont {Khatami}}]{PhysRevE.97.013306}%
  \BibitemOpen
  \bibfield  {author} {\bibinfo {author} {\bibfnamefont {K.}~\bibnamefont {Ch'ng}}, \bibinfo {author} {\bibfnamefont {N.}~\bibnamefont {Vazquez}},\ and\ \bibinfo {author} {\bibfnamefont {E.}~\bibnamefont {Khatami}},\ }\bibfield  {title} {\bibinfo {title} {Unsupervised machine learning account of magnetic transitions in the hubbard model},\ }\href {https://doi.org/10.1103/PhysRevE.97.013306} {\bibfield  {journal} {\bibinfo  {journal} {Phys. Rev. E}\ }\textbf {\bibinfo {volume} {97}},\ \bibinfo {pages} {013306} (\bibinfo {year} {2018})}\BibitemShut {NoStop}%
\bibitem [{\citenamefont {Liu}\ and\ \citenamefont {van Nieuwenburg}(2018)}]{PhysRevLett.120.176401}%
  \BibitemOpen
  \bibfield  {author} {\bibinfo {author} {\bibfnamefont {Y.-H.}\ \bibnamefont {Liu}}\ and\ \bibinfo {author} {\bibfnamefont {E.~P.~L.}\ \bibnamefont {van Nieuwenburg}},\ }\bibfield  {title} {\bibinfo {title} {Discriminative cooperative networks for detecting phase transitions},\ }\href {https://doi.org/10.1103/PhysRevLett.120.176401} {\bibfield  {journal} {\bibinfo  {journal} {Phys. Rev. Lett.}\ }\textbf {\bibinfo {volume} {120}},\ \bibinfo {pages} {176401} (\bibinfo {year} {2018})}\BibitemShut {NoStop}%
\bibitem [{\citenamefont {Huembeli}\ \emph {et~al.}(2018)\citenamefont {Huembeli}, \citenamefont {Dauphin},\ and\ \citenamefont {Wittek}}]{PhysRevB.97.134109}%
  \BibitemOpen
  \bibfield  {author} {\bibinfo {author} {\bibfnamefont {P.}~\bibnamefont {Huembeli}}, \bibinfo {author} {\bibfnamefont {A.}~\bibnamefont {Dauphin}},\ and\ \bibinfo {author} {\bibfnamefont {P.}~\bibnamefont {Wittek}},\ }\bibfield  {title} {\bibinfo {title} {Identifying quantum phase transitions with adversarial neural networks},\ }\href {https://doi.org/10.1103/PhysRevB.97.134109} {\bibfield  {journal} {\bibinfo  {journal} {Phys. Rev. B}\ }\textbf {\bibinfo {volume} {97}},\ \bibinfo {pages} {134109} (\bibinfo {year} {2018})}\BibitemShut {NoStop}%
\bibitem [{\citenamefont {Canabarro}\ \emph {et~al.}(2019)\citenamefont {Canabarro}, \citenamefont {Fanchini}, \citenamefont {Malvezzi}, \citenamefont {Pereira},\ and\ \citenamefont {Chaves}}]{PhysRevB.100.045129}%
  \BibitemOpen
  \bibfield  {author} {\bibinfo {author} {\bibfnamefont {A.}~\bibnamefont {Canabarro}}, \bibinfo {author} {\bibfnamefont {F.~F.}\ \bibnamefont {Fanchini}}, \bibinfo {author} {\bibfnamefont {A.~L.}\ \bibnamefont {Malvezzi}}, \bibinfo {author} {\bibfnamefont {R.}~\bibnamefont {Pereira}},\ and\ \bibinfo {author} {\bibfnamefont {R.}~\bibnamefont {Chaves}},\ }\bibfield  {title} {\bibinfo {title} {Unveiling phase transitions with machine learning},\ }\href {https://doi.org/10.1103/PhysRevB.100.045129} {\bibfield  {journal} {\bibinfo  {journal} {Phys. Rev. B}\ }\textbf {\bibinfo {volume} {100}},\ \bibinfo {pages} {045129} (\bibinfo {year} {2019})}\BibitemShut {NoStop}%
\bibitem [{\citenamefont {Shiina}\ \emph {et~al.}(2020)\citenamefont {Shiina}, \citenamefont {Mori}, \citenamefont {Okabe},\ and\ \citenamefont {Lee}}]{Shiina2020}%
  \BibitemOpen
  \bibfield  {author} {\bibinfo {author} {\bibfnamefont {K.}~\bibnamefont {Shiina}}, \bibinfo {author} {\bibfnamefont {H.}~\bibnamefont {Mori}}, \bibinfo {author} {\bibfnamefont {Y.}~\bibnamefont {Okabe}},\ and\ \bibinfo {author} {\bibfnamefont {H.~K.}\ \bibnamefont {Lee}},\ }\bibfield  {title} {\bibinfo {title} {Machine-learning studies on spin models},\ }\href {https://doi.org/10.1038/s41598-020-58263-5} {\bibfield  {journal} {\bibinfo  {journal} {Scientific Reports}\ }\textbf {\bibinfo {volume} {10}},\ \bibinfo {pages} {2177} (\bibinfo {year} {2020})}\BibitemShut {NoStop}%
\bibitem [{\citenamefont {Tomita}\ \emph {et~al.}(2020)\citenamefont {Tomita}, \citenamefont {Shiina}, \citenamefont {Okabe},\ and\ \citenamefont {Lee}}]{PhysRevE.102.021302}%
  \BibitemOpen
  \bibfield  {author} {\bibinfo {author} {\bibfnamefont {Y.}~\bibnamefont {Tomita}}, \bibinfo {author} {\bibfnamefont {K.}~\bibnamefont {Shiina}}, \bibinfo {author} {\bibfnamefont {Y.}~\bibnamefont {Okabe}},\ and\ \bibinfo {author} {\bibfnamefont {H.~K.}\ \bibnamefont {Lee}},\ }\bibfield  {title} {\bibinfo {title} {Machine-learning study using improved correlation configuration and application to quantum monte carlo simulation},\ }\href {https://doi.org/10.1103/PhysRevE.102.021302} {\bibfield  {journal} {\bibinfo  {journal} {Phys. Rev. E}\ }\textbf {\bibinfo {volume} {102}},\ \bibinfo {pages} {021302} (\bibinfo {year} {2020})}\BibitemShut {NoStop}%
\bibitem [{\citenamefont {Arnold}\ \emph {et~al.}(2021)\citenamefont {Arnold}, \citenamefont {Sch\"afer}, \citenamefont {\ifmmode~\check{Z}\else \v{Z}\fi{}onda},\ and\ \citenamefont {Lode}}]{PhysRevResearch.3.033052}%
  \BibitemOpen
  \bibfield  {author} {\bibinfo {author} {\bibfnamefont {J.}~\bibnamefont {Arnold}}, \bibinfo {author} {\bibfnamefont {F.}~\bibnamefont {Sch\"afer}}, \bibinfo {author} {\bibfnamefont {M.}~\bibnamefont {\ifmmode~\check{Z}\else \v{Z}\fi{}onda}},\ and\ \bibinfo {author} {\bibfnamefont {A.~U.~J.}\ \bibnamefont {Lode}},\ }\bibfield  {title} {\bibinfo {title} {Interpretable and unsupervised phase classification},\ }\href {https://doi.org/10.1103/PhysRevResearch.3.033052} {\bibfield  {journal} {\bibinfo  {journal} {Phys. Rev. Res.}\ }\textbf {\bibinfo {volume} {3}},\ \bibinfo {pages} {033052} (\bibinfo {year} {2021})}\BibitemShut {NoStop}%
\bibitem [{\citenamefont {Tibaldi}\ \emph {et~al.}(2023)\citenamefont {Tibaldi}, \citenamefont {Magnifico}, \citenamefont {Vodola},\ and\ \citenamefont {Ercolessi}}]{10.21468/SciPostPhys.14.1.005}%
  \BibitemOpen
  \bibfield  {author} {\bibinfo {author} {\bibfnamefont {S.}~\bibnamefont {Tibaldi}}, \bibinfo {author} {\bibfnamefont {G.}~\bibnamefont {Magnifico}}, \bibinfo {author} {\bibfnamefont {D.}~\bibnamefont {Vodola}},\ and\ \bibinfo {author} {\bibfnamefont {E.}~\bibnamefont {Ercolessi}},\ }\bibfield  {title} {\bibinfo {title} {{Unsupervised and supervised learning of interacting topological phases from single-particle correlation functions}},\ }\href {https://doi.org/10.21468/SciPostPhys.14.1.005} {\bibfield  {journal} {\bibinfo  {journal} {SciPost Phys.}\ }\textbf {\bibinfo {volume} {14}},\ \bibinfo {pages} {005} (\bibinfo {year} {2023})}\BibitemShut {NoStop}%
\bibitem [{\citenamefont {Chung}\ \emph {et~al.}(2023)\citenamefont {Chung}, \citenamefont {Huang}, \citenamefont {McCulloch},\ and\ \citenamefont {Tsai}}]{chung2023deep}%
  \BibitemOpen
  \bibfield  {author} {\bibinfo {author} {\bibfnamefont {M.-C.}\ \bibnamefont {Chung}}, \bibinfo {author} {\bibfnamefont {G.-Y.}\ \bibnamefont {Huang}}, \bibinfo {author} {\bibfnamefont {I.}~\bibnamefont {McCulloch}},\ and\ \bibinfo {author} {\bibfnamefont {Y.-H.}\ \bibnamefont {Tsai}},\ }\href@noop {} {\bibinfo {title} {Deep learning of phase transitions for quantum spin chains from correlation aspects}} (\bibinfo {year} {2023}),\ \Eprint {https://arxiv.org/abs/2301.06669} {arXiv:2301.06669 [cond-mat.stat-mech]} \BibitemShut {NoStop}%
\bibitem [{\citenamefont {Hasan}\ and\ \citenamefont {Kane}(2010)}]{RevModPhys.82.3045}%
  \BibitemOpen
  \bibfield  {author} {\bibinfo {author} {\bibfnamefont {M.~Z.}\ \bibnamefont {Hasan}}\ and\ \bibinfo {author} {\bibfnamefont {C.~L.}\ \bibnamefont {Kane}},\ }\bibfield  {title} {\bibinfo {title} {Colloquium: Topological insulators},\ }\href {https://doi.org/10.1103/RevModPhys.82.3045} {\bibfield  {journal} {\bibinfo  {journal} {Rev. Mod. Phys.}\ }\textbf {\bibinfo {volume} {82}},\ \bibinfo {pages} {3045} (\bibinfo {year} {2010})}\BibitemShut {NoStop}%
\bibitem [{\citenamefont {Zhang}\ and\ \citenamefont {Kim}(2017)}]{PhysRevLett.118.216401}%
  \BibitemOpen
  \bibfield  {author} {\bibinfo {author} {\bibfnamefont {Y.}~\bibnamefont {Zhang}}\ and\ \bibinfo {author} {\bibfnamefont {E.-A.}\ \bibnamefont {Kim}},\ }\bibfield  {title} {\bibinfo {title} {Quantum loop topography for machine learning},\ }\href {https://doi.org/10.1103/PhysRevLett.118.216401} {\bibfield  {journal} {\bibinfo  {journal} {Phys. Rev. Lett.}\ }\textbf {\bibinfo {volume} {118}},\ \bibinfo {pages} {216401} (\bibinfo {year} {2017})}\BibitemShut {NoStop}%
\bibitem [{\citenamefont {Zhang}\ \emph {et~al.}(2017)\citenamefont {Zhang}, \citenamefont {Melko},\ and\ \citenamefont {Kim}}]{PhysRevB.96.245119}%
  \BibitemOpen
  \bibfield  {author} {\bibinfo {author} {\bibfnamefont {Y.}~\bibnamefont {Zhang}}, \bibinfo {author} {\bibfnamefont {R.~G.}\ \bibnamefont {Melko}},\ and\ \bibinfo {author} {\bibfnamefont {E.-A.}\ \bibnamefont {Kim}},\ }\bibfield  {title} {\bibinfo {title} {Machine learning ${\mathbb{z}}_{2}$ quantum spin liquids with quasiparticle statistics},\ }\href {https://doi.org/10.1103/PhysRevB.96.245119} {\bibfield  {journal} {\bibinfo  {journal} {Phys. Rev. B}\ }\textbf {\bibinfo {volume} {96}},\ \bibinfo {pages} {245119} (\bibinfo {year} {2017})}\BibitemShut {NoStop}%
\bibitem [{\citenamefont {Deng}\ \emph {et~al.}(2017)\citenamefont {Deng}, \citenamefont {Li},\ and\ \citenamefont {Das~Sarma}}]{PhysRevB.96.195145}%
  \BibitemOpen
  \bibfield  {author} {\bibinfo {author} {\bibfnamefont {D.-L.}\ \bibnamefont {Deng}}, \bibinfo {author} {\bibfnamefont {X.}~\bibnamefont {Li}},\ and\ \bibinfo {author} {\bibfnamefont {S.}~\bibnamefont {Das~Sarma}},\ }\bibfield  {title} {\bibinfo {title} {Machine learning topological states},\ }\href {https://doi.org/10.1103/PhysRevB.96.195145} {\bibfield  {journal} {\bibinfo  {journal} {Phys. Rev. B}\ }\textbf {\bibinfo {volume} {96}},\ \bibinfo {pages} {195145} (\bibinfo {year} {2017})}\BibitemShut {NoStop}%
\bibitem [{\citenamefont {Zhang}\ \emph {et~al.}(2018)\citenamefont {Zhang}, \citenamefont {Shen},\ and\ \citenamefont {Zhai}}]{PhysRevLett.120.066401}%
  \BibitemOpen
  \bibfield  {author} {\bibinfo {author} {\bibfnamefont {P.}~\bibnamefont {Zhang}}, \bibinfo {author} {\bibfnamefont {H.}~\bibnamefont {Shen}},\ and\ \bibinfo {author} {\bibfnamefont {H.}~\bibnamefont {Zhai}},\ }\bibfield  {title} {\bibinfo {title} {Machine learning topological invariants with neural networks},\ }\href {https://doi.org/10.1103/PhysRevLett.120.066401} {\bibfield  {journal} {\bibinfo  {journal} {Phys. Rev. Lett.}\ }\textbf {\bibinfo {volume} {120}},\ \bibinfo {pages} {066401} (\bibinfo {year} {2018})}\BibitemShut {NoStop}%
\bibitem [{\citenamefont {Sun}\ \emph {et~al.}(2018)\citenamefont {Sun}, \citenamefont {Yi}, \citenamefont {Zhang}, \citenamefont {Shen},\ and\ \citenamefont {Zhai}}]{PhysRevB.98.085402}%
  \BibitemOpen
  \bibfield  {author} {\bibinfo {author} {\bibfnamefont {N.}~\bibnamefont {Sun}}, \bibinfo {author} {\bibfnamefont {J.}~\bibnamefont {Yi}}, \bibinfo {author} {\bibfnamefont {P.}~\bibnamefont {Zhang}}, \bibinfo {author} {\bibfnamefont {H.}~\bibnamefont {Shen}},\ and\ \bibinfo {author} {\bibfnamefont {H.}~\bibnamefont {Zhai}},\ }\bibfield  {title} {\bibinfo {title} {Deep learning topological invariants of band insulators},\ }\href {https://doi.org/10.1103/PhysRevB.98.085402} {\bibfield  {journal} {\bibinfo  {journal} {Phys. Rev. B}\ }\textbf {\bibinfo {volume} {98}},\ \bibinfo {pages} {085402} (\bibinfo {year} {2018})}\BibitemShut {NoStop}%
\bibitem [{\citenamefont {Carvalho}\ \emph {et~al.}(2018)\citenamefont {Carvalho}, \citenamefont {Garc\'{\i}a-Mart\'{\i}nez}, \citenamefont {Lado},\ and\ \citenamefont {Fern\'andez-Rossier}}]{PhysRevB.97.115453}%
  \BibitemOpen
  \bibfield  {author} {\bibinfo {author} {\bibfnamefont {D.}~\bibnamefont {Carvalho}}, \bibinfo {author} {\bibfnamefont {N.~A.}\ \bibnamefont {Garc\'{\i}a-Mart\'{\i}nez}}, \bibinfo {author} {\bibfnamefont {J.~L.}\ \bibnamefont {Lado}},\ and\ \bibinfo {author} {\bibfnamefont {J.}~\bibnamefont {Fern\'andez-Rossier}},\ }\bibfield  {title} {\bibinfo {title} {Real-space mapping of topological invariants using artificial neural networks},\ }\href {https://doi.org/10.1103/PhysRevB.97.115453} {\bibfield  {journal} {\bibinfo  {journal} {Phys. Rev. B}\ }\textbf {\bibinfo {volume} {97}},\ \bibinfo {pages} {115453} (\bibinfo {year} {2018})}\BibitemShut {NoStop}%
\bibitem [{\citenamefont {Rodriguez-Nieva}\ and\ \citenamefont {Scheurer}(2019)}]{Rodriguez-Nieva2019}%
  \BibitemOpen
  \bibfield  {author} {\bibinfo {author} {\bibfnamefont {J.~F.}\ \bibnamefont {Rodriguez-Nieva}}\ and\ \bibinfo {author} {\bibfnamefont {M.~S.}\ \bibnamefont {Scheurer}},\ }\bibfield  {title} {\bibinfo {title} {Identifying topological order through unsupervised machine learning},\ }\href {https://doi.org/10.1038/s41567-019-0512-x} {\bibfield  {journal} {\bibinfo  {journal} {Nature Physics}\ }\textbf {\bibinfo {volume} {15}},\ \bibinfo {pages} {790} (\bibinfo {year} {2019})}\BibitemShut {NoStop}%
\bibitem [{\citenamefont {Ming}\ \emph {et~al.}(2019)\citenamefont {Ming}, \citenamefont {Lin}, \citenamefont {Bartlett},\ and\ \citenamefont {Zhang}}]{Ming2019}%
  \BibitemOpen
  \bibfield  {author} {\bibinfo {author} {\bibfnamefont {Y.}~\bibnamefont {Ming}}, \bibinfo {author} {\bibfnamefont {C.-T.}\ \bibnamefont {Lin}}, \bibinfo {author} {\bibfnamefont {S.~D.}\ \bibnamefont {Bartlett}},\ and\ \bibinfo {author} {\bibfnamefont {W.-W.}\ \bibnamefont {Zhang}},\ }\bibfield  {title} {\bibinfo {title} {Quantum topology identification with deep neural networks and quantum walks},\ }\href {https://doi.org/10.1038/s41524-019-0224-x} {\bibfield  {journal} {\bibinfo  {journal} {npj Computational Materials}\ }\textbf {\bibinfo {volume} {5}},\ \bibinfo {pages} {88} (\bibinfo {year} {2019})}\BibitemShut {NoStop}%
\bibitem [{\citenamefont {Caio}\ \emph {et~al.}(2019)\citenamefont {Caio}, \citenamefont {Caccin}, \citenamefont {Baireuther}, \citenamefont {Hyart},\ and\ \citenamefont {Fruchart}}]{https://doi.org/10.48550/arxiv.1901.03346}%
  \BibitemOpen
  \bibfield  {author} {\bibinfo {author} {\bibfnamefont {M.~D.}\ \bibnamefont {Caio}}, \bibinfo {author} {\bibfnamefont {M.}~\bibnamefont {Caccin}}, \bibinfo {author} {\bibfnamefont {P.}~\bibnamefont {Baireuther}}, \bibinfo {author} {\bibfnamefont {T.}~\bibnamefont {Hyart}},\ and\ \bibinfo {author} {\bibfnamefont {M.}~\bibnamefont {Fruchart}},\ }\href {https://doi.org/10.48550/ARXIV.1901.03346} {\bibinfo {title} {Machine learning assisted measurement of local topological invariants}} (\bibinfo {year} {2019})\BibitemShut {NoStop}%
\bibitem [{\citenamefont {Scheurer}\ and\ \citenamefont {Slager}(2020)}]{PhysRevLett.124.226401}%
  \BibitemOpen
  \bibfield  {author} {\bibinfo {author} {\bibfnamefont {M.~S.}\ \bibnamefont {Scheurer}}\ and\ \bibinfo {author} {\bibfnamefont {R.-J.}\ \bibnamefont {Slager}},\ }\bibfield  {title} {\bibinfo {title} {Unsupervised machine learning and band topology},\ }\href {https://doi.org/10.1103/PhysRevLett.124.226401} {\bibfield  {journal} {\bibinfo  {journal} {Phys. Rev. Lett.}\ }\textbf {\bibinfo {volume} {124}},\ \bibinfo {pages} {226401} (\bibinfo {year} {2020})}\BibitemShut {NoStop}%
\bibitem [{\citenamefont {Greplova}\ \emph {et~al.}(2020)\citenamefont {Greplova}, \citenamefont {Valenti}, \citenamefont {Boschung}, \citenamefont {Schäfer}, \citenamefont {Lörch},\ and\ \citenamefont {Huber}}]{Greplova_2020}%
  \BibitemOpen
  \bibfield  {author} {\bibinfo {author} {\bibfnamefont {E.}~\bibnamefont {Greplova}}, \bibinfo {author} {\bibfnamefont {A.}~\bibnamefont {Valenti}}, \bibinfo {author} {\bibfnamefont {G.}~\bibnamefont {Boschung}}, \bibinfo {author} {\bibfnamefont {F.}~\bibnamefont {Schäfer}}, \bibinfo {author} {\bibfnamefont {N.}~\bibnamefont {Lörch}},\ and\ \bibinfo {author} {\bibfnamefont {S.~D.}\ \bibnamefont {Huber}},\ }\bibfield  {title} {\bibinfo {title} {Unsupervised identification of topological phase transitions using predictive models},\ }\href {https://doi.org/10.1088/1367-2630/ab7771} {\bibfield  {journal} {\bibinfo  {journal} {New Journal of Physics}\ }\textbf {\bibinfo {volume} {22}},\ \bibinfo {pages} {045003} (\bibinfo {year} {2020})}\BibitemShut {NoStop}%
\bibitem [{\citenamefont {Tsai}\ \emph {et~al.}(2020)\citenamefont {Tsai}, \citenamefont {Yu}, \citenamefont {Hsu},\ and\ \citenamefont {Chung}}]{PhysRevB.102.054512}%
  \BibitemOpen
  \bibfield  {author} {\bibinfo {author} {\bibfnamefont {Y.-H.}\ \bibnamefont {Tsai}}, \bibinfo {author} {\bibfnamefont {M.-Z.}\ \bibnamefont {Yu}}, \bibinfo {author} {\bibfnamefont {Y.-H.}\ \bibnamefont {Hsu}},\ and\ \bibinfo {author} {\bibfnamefont {M.-C.}\ \bibnamefont {Chung}},\ }\bibfield  {title} {\bibinfo {title} {Deep learning of topological phase transitions from entanglement aspects},\ }\href {https://doi.org/10.1103/PhysRevB.102.054512} {\bibfield  {journal} {\bibinfo  {journal} {Phys. Rev. B}\ }\textbf {\bibinfo {volume} {102}},\ \bibinfo {pages} {054512} (\bibinfo {year} {2020})}\BibitemShut {NoStop}%
\bibitem [{\citenamefont {Zhang}\ \emph {et~al.}(2021)\citenamefont {Zhang}, \citenamefont {Tang}, \citenamefont {Huang}, \citenamefont {Zhang}, \citenamefont {Huang},\ and\ \citenamefont {Zhang}}]{PhysRevA.103.012419}%
  \BibitemOpen
  \bibfield  {author} {\bibinfo {author} {\bibfnamefont {L.-F.}\ \bibnamefont {Zhang}}, \bibinfo {author} {\bibfnamefont {L.-Z.}\ \bibnamefont {Tang}}, \bibinfo {author} {\bibfnamefont {Z.-H.}\ \bibnamefont {Huang}}, \bibinfo {author} {\bibfnamefont {G.-Q.}\ \bibnamefont {Zhang}}, \bibinfo {author} {\bibfnamefont {W.}~\bibnamefont {Huang}},\ and\ \bibinfo {author} {\bibfnamefont {D.-W.}\ \bibnamefont {Zhang}},\ }\bibfield  {title} {\bibinfo {title} {Machine learning topological invariants of non-hermitian systems},\ }\href {https://doi.org/10.1103/PhysRevA.103.012419} {\bibfield  {journal} {\bibinfo  {journal} {Phys. Rev. A}\ }\textbf {\bibinfo {volume} {103}},\ \bibinfo {pages} {012419} (\bibinfo {year} {2021})}\BibitemShut {NoStop}%
\bibitem [{\citenamefont {Chung}\ \emph {et~al.}(2021)\citenamefont {Chung}, \citenamefont {Cheng}, \citenamefont {Huang},\ and\ \citenamefont {Tsai}}]{PhysRevB.104.024506}%
  \BibitemOpen
  \bibfield  {author} {\bibinfo {author} {\bibfnamefont {M.-C.}\ \bibnamefont {Chung}}, \bibinfo {author} {\bibfnamefont {T.-P.}\ \bibnamefont {Cheng}}, \bibinfo {author} {\bibfnamefont {G.-Y.}\ \bibnamefont {Huang}},\ and\ \bibinfo {author} {\bibfnamefont {Y.-H.}\ \bibnamefont {Tsai}},\ }\bibfield  {title} {\bibinfo {title} {Deep learning of topological phase transitions from the point of view of entanglement for two-dimensional chiral $p$-wave superconductors},\ }\href {https://doi.org/10.1103/PhysRevB.104.024506} {\bibfield  {journal} {\bibinfo  {journal} {Phys. Rev. B}\ }\textbf {\bibinfo {volume} {104}},\ \bibinfo {pages} {024506} (\bibinfo {year} {2021})}\BibitemShut {NoStop}%
\bibitem [{\citenamefont {Tsai}\ \emph {et~al.}(2021)\citenamefont {Tsai}, \citenamefont {Chiu}, \citenamefont {Lai}, \citenamefont {Su}, \citenamefont {Yang}, \citenamefont {Cheng}, \citenamefont {Huang},\ and\ \citenamefont {Chung}}]{PhysRevB.104.165108}%
  \BibitemOpen
  \bibfield  {author} {\bibinfo {author} {\bibfnamefont {Y.-H.}\ \bibnamefont {Tsai}}, \bibinfo {author} {\bibfnamefont {K.-F.}\ \bibnamefont {Chiu}}, \bibinfo {author} {\bibfnamefont {Y.-C.}\ \bibnamefont {Lai}}, \bibinfo {author} {\bibfnamefont {K.-J.}\ \bibnamefont {Su}}, \bibinfo {author} {\bibfnamefont {T.-P.}\ \bibnamefont {Yang}}, \bibinfo {author} {\bibfnamefont {T.-P.}\ \bibnamefont {Cheng}}, \bibinfo {author} {\bibfnamefont {G.-Y.}\ \bibnamefont {Huang}},\ and\ \bibinfo {author} {\bibfnamefont {M.-C.}\ \bibnamefont {Chung}},\ }\bibfield  {title} {\bibinfo {title} {Deep learning of topological phase transitions from entanglement aspects: An unsupervised way},\ }\href {https://doi.org/10.1103/PhysRevB.104.165108} {\bibfield  {journal} {\bibinfo  {journal} {Phys. Rev. B}\ }\textbf {\bibinfo {volume} {104}},\ \bibinfo {pages} {165108} (\bibinfo {year} {2021})}\BibitemShut {NoStop}%
\bibitem [{\citenamefont {Molignini}\ \emph {et~al.}(2021)\citenamefont {Molignini}, \citenamefont {Zegarra}, \citenamefont {van Nieuwenburg}, \citenamefont {Chitra},\ and\ \citenamefont {Chen}}]{10.21468/SciPostPhys.11.3.073}%
  \BibitemOpen
  \bibfield  {author} {\bibinfo {author} {\bibfnamefont {P.}~\bibnamefont {Molignini}}, \bibinfo {author} {\bibfnamefont {A.}~\bibnamefont {Zegarra}}, \bibinfo {author} {\bibfnamefont {E.}~\bibnamefont {van Nieuwenburg}}, \bibinfo {author} {\bibfnamefont {R.}~\bibnamefont {Chitra}},\ and\ \bibinfo {author} {\bibfnamefont {W.}~\bibnamefont {Chen}},\ }\bibfield  {title} {\bibinfo {title} {{A supervised learning algorithm for interacting topological insulators based on local curvature}},\ }\href {https://doi.org/10.21468/SciPostPhys.11.3.073} {\bibfield  {journal} {\bibinfo  {journal} {SciPost Phys.}\ }\textbf {\bibinfo {volume} {11}},\ \bibinfo {pages} {073} (\bibinfo {year} {2021})}\BibitemShut {NoStop}%
\bibitem [{\citenamefont {Chiu}\ \emph {et~al.}(2013)\citenamefont {Chiu}, \citenamefont {Yao},\ and\ \citenamefont {Ryu}}]{PhysRevB.88.075142}%
  \BibitemOpen
  \bibfield  {author} {\bibinfo {author} {\bibfnamefont {C.-K.}\ \bibnamefont {Chiu}}, \bibinfo {author} {\bibfnamefont {H.}~\bibnamefont {Yao}},\ and\ \bibinfo {author} {\bibfnamefont {S.}~\bibnamefont {Ryu}},\ }\bibfield  {title} {\bibinfo {title} {Classification of topological insulators and superconductors in the presence of reflection symmetry},\ }\href {https://doi.org/10.1103/PhysRevB.88.075142} {\bibfield  {journal} {\bibinfo  {journal} {Phys. Rev. B}\ }\textbf {\bibinfo {volume} {88}},\ \bibinfo {pages} {075142} (\bibinfo {year} {2013})}\BibitemShut {NoStop}%
\bibitem [{\citenamefont {Ghosh}\ and\ \citenamefont {Grosfeld}(2021)}]{10.21468/SciPostPhys.10.3.059}%
  \BibitemOpen
  \bibfield  {author} {\bibinfo {author} {\bibfnamefont {A.}~\bibnamefont {Ghosh}}\ and\ \bibinfo {author} {\bibfnamefont {E.}~\bibnamefont {Grosfeld}},\ }\bibfield  {title} {\bibinfo {title} {{Weak topological insulating phases of hard-core-bosons on the honeycomb lattice}},\ }\href {https://doi.org/10.21468/SciPostPhys.10.3.059} {\bibfield  {journal} {\bibinfo  {journal} {SciPost Phys.}\ }\textbf {\bibinfo {volume} {10}},\ \bibinfo {pages} {059} (\bibinfo {year} {2021})}\BibitemShut {NoStop}%
\bibitem [{\citenamefont {Dong}\ \emph {et~al.}(2019)\citenamefont {Dong}, \citenamefont {Pollmann},\ and\ \citenamefont {Zhang}}]{PhysRevB.99.121104}%
  \BibitemOpen
  \bibfield  {author} {\bibinfo {author} {\bibfnamefont {X.-Y.}\ \bibnamefont {Dong}}, \bibinfo {author} {\bibfnamefont {F.}~\bibnamefont {Pollmann}},\ and\ \bibinfo {author} {\bibfnamefont {X.-F.}\ \bibnamefont {Zhang}},\ }\bibfield  {title} {\bibinfo {title} {Machine learning of quantum phase transitions},\ }\href {https://doi.org/10.1103/PhysRevB.99.121104} {\bibfield  {journal} {\bibinfo  {journal} {Phys. Rev. B}\ }\textbf {\bibinfo {volume} {99}},\ \bibinfo {pages} {121104} (\bibinfo {year} {2019})}\BibitemShut {NoStop}%
\bibitem [{\citenamefont {Carleo}\ and\ \citenamefont {Troyer}(2017)}]{doi:10.1126/science.aag2302}%
  \BibitemOpen
  \bibfield  {author} {\bibinfo {author} {\bibfnamefont {G.}~\bibnamefont {Carleo}}\ and\ \bibinfo {author} {\bibfnamefont {M.}~\bibnamefont {Troyer}},\ }\bibfield  {title} {\bibinfo {title} {Solving the quantum many-body problem with artificial neural networks},\ }\href {https://doi.org/10.1126/science.aag2302} {\bibfield  {journal} {\bibinfo  {journal} {Science}\ }\textbf {\bibinfo {volume} {355}},\ \bibinfo {pages} {602} (\bibinfo {year} {2017})},\ \Eprint {https://arxiv.org/abs/https://www.science.org/doi/pdf/10.1126/science.aag2302} {https://www.science.org/doi/pdf/10.1126/science.aag2302} \BibitemShut {NoStop}%
\bibitem [{\citenamefont {Schindler}\ \emph {et~al.}(2017)\citenamefont {Schindler}, \citenamefont {Regnault},\ and\ \citenamefont {Neupert}}]{PhysRevB.95.245134}%
  \BibitemOpen
  \bibfield  {author} {\bibinfo {author} {\bibfnamefont {F.}~\bibnamefont {Schindler}}, \bibinfo {author} {\bibfnamefont {N.}~\bibnamefont {Regnault}},\ and\ \bibinfo {author} {\bibfnamefont {T.}~\bibnamefont {Neupert}},\ }\bibfield  {title} {\bibinfo {title} {Probing many-body localization with neural networks},\ }\href {https://doi.org/10.1103/PhysRevB.95.245134} {\bibfield  {journal} {\bibinfo  {journal} {Phys. Rev. B}\ }\textbf {\bibinfo {volume} {95}},\ \bibinfo {pages} {245134} (\bibinfo {year} {2017})}\BibitemShut {NoStop}%
\bibitem [{\citenamefont {Torlai}\ \emph {et~al.}(2018)\citenamefont {Torlai}, \citenamefont {Mazzola}, \citenamefont {Carrasquilla}, \citenamefont {Troyer}, \citenamefont {Melko},\ and\ \citenamefont {Carleo}}]{Torlai2018}%
  \BibitemOpen
  \bibfield  {author} {\bibinfo {author} {\bibfnamefont {G.}~\bibnamefont {Torlai}}, \bibinfo {author} {\bibfnamefont {G.}~\bibnamefont {Mazzola}}, \bibinfo {author} {\bibfnamefont {J.}~\bibnamefont {Carrasquilla}}, \bibinfo {author} {\bibfnamefont {M.}~\bibnamefont {Troyer}}, \bibinfo {author} {\bibfnamefont {R.}~\bibnamefont {Melko}},\ and\ \bibinfo {author} {\bibfnamefont {G.}~\bibnamefont {Carleo}},\ }\bibfield  {title} {\bibinfo {title} {Neural-network quantum state tomography},\ }\href {https://doi.org/10.1038/s41567-018-0048-5} {\bibfield  {journal} {\bibinfo  {journal} {Nature Physics}\ }\textbf {\bibinfo {volume} {14}},\ \bibinfo {pages} {447} (\bibinfo {year} {2018})}\BibitemShut {NoStop}%
\bibitem [{\citenamefont {Liang}\ \emph {et~al.}(2018)\citenamefont {Liang}, \citenamefont {Liu}, \citenamefont {Lin}, \citenamefont {Guo}, \citenamefont {Zhang},\ and\ \citenamefont {He}}]{PhysRevB.98.104426}%
  \BibitemOpen
  \bibfield  {author} {\bibinfo {author} {\bibfnamefont {X.}~\bibnamefont {Liang}}, \bibinfo {author} {\bibfnamefont {W.-Y.}\ \bibnamefont {Liu}}, \bibinfo {author} {\bibfnamefont {P.-Z.}\ \bibnamefont {Lin}}, \bibinfo {author} {\bibfnamefont {G.-C.}\ \bibnamefont {Guo}}, \bibinfo {author} {\bibfnamefont {Y.-S.}\ \bibnamefont {Zhang}},\ and\ \bibinfo {author} {\bibfnamefont {L.}~\bibnamefont {He}},\ }\bibfield  {title} {\bibinfo {title} {Solving frustrated quantum many-particle models with convolutional neural networks},\ }\href {https://doi.org/10.1103/PhysRevB.98.104426} {\bibfield  {journal} {\bibinfo  {journal} {Phys. Rev. B}\ }\textbf {\bibinfo {volume} {98}},\ \bibinfo {pages} {104426} (\bibinfo {year} {2018})}\BibitemShut {NoStop}%
\bibitem [{\citenamefont {Choo}\ \emph {et~al.}(2018)\citenamefont {Choo}, \citenamefont {Carleo}, \citenamefont {Regnault},\ and\ \citenamefont {Neupert}}]{PhysRevLett.121.167204}%
  \BibitemOpen
  \bibfield  {author} {\bibinfo {author} {\bibfnamefont {K.}~\bibnamefont {Choo}}, \bibinfo {author} {\bibfnamefont {G.}~\bibnamefont {Carleo}}, \bibinfo {author} {\bibfnamefont {N.}~\bibnamefont {Regnault}},\ and\ \bibinfo {author} {\bibfnamefont {T.}~\bibnamefont {Neupert}},\ }\bibfield  {title} {\bibinfo {title} {Symmetries and many-body excitations with neural-network quantum states},\ }\href {https://doi.org/10.1103/PhysRevLett.121.167204} {\bibfield  {journal} {\bibinfo  {journal} {Phys. Rev. Lett.}\ }\textbf {\bibinfo {volume} {121}},\ \bibinfo {pages} {167204} (\bibinfo {year} {2018})}\BibitemShut {NoStop}%
\bibitem [{\citenamefont {Th\'eveniaut}\ and\ \citenamefont {Alet}(2019)}]{PhysRevB.100.224202}%
  \BibitemOpen
  \bibfield  {author} {\bibinfo {author} {\bibfnamefont {H.}~\bibnamefont {Th\'eveniaut}}\ and\ \bibinfo {author} {\bibfnamefont {F.}~\bibnamefont {Alet}},\ }\bibfield  {title} {\bibinfo {title} {Neural network setups for a precise detection of the many-body localization transition: Finite-size scaling and limitations},\ }\href {https://doi.org/10.1103/PhysRevB.100.224202} {\bibfield  {journal} {\bibinfo  {journal} {Phys. Rev. B}\ }\textbf {\bibinfo {volume} {100}},\ \bibinfo {pages} {224202} (\bibinfo {year} {2019})}\BibitemShut {NoStop}%
\bibitem [{\citenamefont {Raghu}\ \emph {et~al.}(2008)\citenamefont {Raghu}, \citenamefont {Qi}, \citenamefont {Honerkamp},\ and\ \citenamefont {Zhang}}]{PhysRevLett.100.156401}%
  \BibitemOpen
  \bibfield  {author} {\bibinfo {author} {\bibfnamefont {S.}~\bibnamefont {Raghu}}, \bibinfo {author} {\bibfnamefont {X.-L.}\ \bibnamefont {Qi}}, \bibinfo {author} {\bibfnamefont {C.}~\bibnamefont {Honerkamp}},\ and\ \bibinfo {author} {\bibfnamefont {S.-C.}\ \bibnamefont {Zhang}},\ }\bibfield  {title} {\bibinfo {title} {Topological mott insulators},\ }\href {https://doi.org/10.1103/PhysRevLett.100.156401} {\bibfield  {journal} {\bibinfo  {journal} {Phys. Rev. Lett.}\ }\textbf {\bibinfo {volume} {100}},\ \bibinfo {pages} {156401} (\bibinfo {year} {2008})}\BibitemShut {NoStop}%
\bibitem [{\citenamefont {Weeks}\ and\ \citenamefont {Franz}(2010)}]{PhysRevB.81.085105}%
  \BibitemOpen
  \bibfield  {author} {\bibinfo {author} {\bibfnamefont {C.}~\bibnamefont {Weeks}}\ and\ \bibinfo {author} {\bibfnamefont {M.}~\bibnamefont {Franz}},\ }\bibfield  {title} {\bibinfo {title} {Interaction-driven instabilities of a dirac semimetal},\ }\href {https://doi.org/10.1103/PhysRevB.81.085105} {\bibfield  {journal} {\bibinfo  {journal} {Phys. Rev. B}\ }\textbf {\bibinfo {volume} {81}},\ \bibinfo {pages} {085105} (\bibinfo {year} {2010})}\BibitemShut {NoStop}%
\bibitem [{\citenamefont {Pesin}\ and\ \citenamefont {Balents}(2010)}]{Pesin2010}%
  \BibitemOpen
  \bibfield  {author} {\bibinfo {author} {\bibfnamefont {D.}~\bibnamefont {Pesin}}\ and\ \bibinfo {author} {\bibfnamefont {L.}~\bibnamefont {Balents}},\ }\bibfield  {title} {\bibinfo {title} {Mott physics and band topology in materials with strong spin--orbit interaction},\ }\href {https://doi.org/10.1038/nphys1606} {\bibfield  {journal} {\bibinfo  {journal} {Nature Physics}\ }\textbf {\bibinfo {volume} {6}},\ \bibinfo {pages} {376} (\bibinfo {year} {2010})}\BibitemShut {NoStop}%
\bibitem [{\citenamefont {Hohenadler}\ \emph {et~al.}(2011)\citenamefont {Hohenadler}, \citenamefont {Lang},\ and\ \citenamefont {Assaad}}]{PhysRevLett.106.100403}%
  \BibitemOpen
  \bibfield  {author} {\bibinfo {author} {\bibfnamefont {M.}~\bibnamefont {Hohenadler}}, \bibinfo {author} {\bibfnamefont {T.~C.}\ \bibnamefont {Lang}},\ and\ \bibinfo {author} {\bibfnamefont {F.~F.}\ \bibnamefont {Assaad}},\ }\bibfield  {title} {\bibinfo {title} {Correlation effects in quantum spin-hall insulators: A quantum monte carlo study},\ }\href {https://doi.org/10.1103/PhysRevLett.106.100403} {\bibfield  {journal} {\bibinfo  {journal} {Phys. Rev. Lett.}\ }\textbf {\bibinfo {volume} {106}},\ \bibinfo {pages} {100403} (\bibinfo {year} {2011})}\BibitemShut {NoStop}%
\bibitem [{\citenamefont {Castro}\ \emph {et~al.}(2011)\citenamefont {Castro}, \citenamefont {Grushin}, \citenamefont {Valenzuela}, \citenamefont {Vozmediano}, \citenamefont {Cortijo},\ and\ \citenamefont {de~Juan}}]{PhysRevLett.107.106402}%
  \BibitemOpen
  \bibfield  {author} {\bibinfo {author} {\bibfnamefont {E.~V.}\ \bibnamefont {Castro}}, \bibinfo {author} {\bibfnamefont {A.~G.}\ \bibnamefont {Grushin}}, \bibinfo {author} {\bibfnamefont {B.}~\bibnamefont {Valenzuela}}, \bibinfo {author} {\bibfnamefont {M.~A.~H.}\ \bibnamefont {Vozmediano}}, \bibinfo {author} {\bibfnamefont {A.}~\bibnamefont {Cortijo}},\ and\ \bibinfo {author} {\bibfnamefont {F.}~\bibnamefont {de~Juan}},\ }\bibfield  {title} {\bibinfo {title} {Topological fermi liquids from coulomb interactions in the doped honeycomb lattice},\ }\href {https://doi.org/10.1103/PhysRevLett.107.106402} {\bibfield  {journal} {\bibinfo  {journal} {Phys. Rev. Lett.}\ }\textbf {\bibinfo {volume} {107}},\ \bibinfo {pages} {106402} (\bibinfo {year} {2011})}\BibitemShut {NoStop}%
\bibitem [{\citenamefont {Lee}(2011)}]{PhysRevLett.107.166806}%
  \BibitemOpen
  \bibfield  {author} {\bibinfo {author} {\bibfnamefont {D.-H.}\ \bibnamefont {Lee}},\ }\bibfield  {title} {\bibinfo {title} {Effects of interaction on quantum spin hall insulators},\ }\href {https://doi.org/10.1103/PhysRevLett.107.166806} {\bibfield  {journal} {\bibinfo  {journal} {Phys. Rev. Lett.}\ }\textbf {\bibinfo {volume} {107}},\ \bibinfo {pages} {166806} (\bibinfo {year} {2011})}\BibitemShut {NoStop}%
\bibitem [{\citenamefont {Ara\'ujo}\ \emph {et~al.}(2013)\citenamefont {Ara\'ujo}, \citenamefont {Castro},\ and\ \citenamefont {Sacramento}}]{PhysRevB.87.085109}%
  \BibitemOpen
  \bibfield  {author} {\bibinfo {author} {\bibfnamefont {M.~A.~N.}\ \bibnamefont {Ara\'ujo}}, \bibinfo {author} {\bibfnamefont {E.~V.}\ \bibnamefont {Castro}},\ and\ \bibinfo {author} {\bibfnamefont {P.~D.}\ \bibnamefont {Sacramento}},\ }\bibfield  {title} {\bibinfo {title} {Change of an insulator's topological properties by a hubbard interaction},\ }\href {https://doi.org/10.1103/PhysRevB.87.085109} {\bibfield  {journal} {\bibinfo  {journal} {Phys. Rev. B}\ }\textbf {\bibinfo {volume} {87}},\ \bibinfo {pages} {085109} (\bibinfo {year} {2013})}\BibitemShut {NoStop}%
\bibitem [{\citenamefont {Grushin}\ \emph {et~al.}(2013)\citenamefont {Grushin}, \citenamefont {Castro}, \citenamefont {Cortijo}, \citenamefont {de~Juan}, \citenamefont {Vozmediano},\ and\ \citenamefont {Valenzuela}}]{PhysRevB.87.085136}%
  \BibitemOpen
  \bibfield  {author} {\bibinfo {author} {\bibfnamefont {A.~G.}\ \bibnamefont {Grushin}}, \bibinfo {author} {\bibfnamefont {E.~V.}\ \bibnamefont {Castro}}, \bibinfo {author} {\bibfnamefont {A.}~\bibnamefont {Cortijo}}, \bibinfo {author} {\bibfnamefont {F.}~\bibnamefont {de~Juan}}, \bibinfo {author} {\bibfnamefont {M.~A.~H.}\ \bibnamefont {Vozmediano}},\ and\ \bibinfo {author} {\bibfnamefont {B.}~\bibnamefont {Valenzuela}},\ }\bibfield  {title} {\bibinfo {title} {Charge instabilities and topological phases in the extended hubbard model on the honeycomb lattice with enlarged unit cell},\ }\href {https://doi.org/10.1103/PhysRevB.87.085136} {\bibfield  {journal} {\bibinfo  {journal} {Phys. Rev. B}\ }\textbf {\bibinfo {volume} {87}},\ \bibinfo {pages} {085136} (\bibinfo {year} {2013})}\BibitemShut {NoStop}%
\bibitem [{\citenamefont {Garc\'{\i}a-Mart\'{\i}nez}\ \emph {et~al.}(2013)\citenamefont {Garc\'{\i}a-Mart\'{\i}nez}, \citenamefont {Grushin}, \citenamefont {Neupert}, \citenamefont {Valenzuela},\ and\ \citenamefont {Castro}}]{PhysRevB.88.245123}%
  \BibitemOpen
  \bibfield  {author} {\bibinfo {author} {\bibfnamefont {N.~A.}\ \bibnamefont {Garc\'{\i}a-Mart\'{\i}nez}}, \bibinfo {author} {\bibfnamefont {A.~G.}\ \bibnamefont {Grushin}}, \bibinfo {author} {\bibfnamefont {T.}~\bibnamefont {Neupert}}, \bibinfo {author} {\bibfnamefont {B.}~\bibnamefont {Valenzuela}},\ and\ \bibinfo {author} {\bibfnamefont {E.~V.}\ \bibnamefont {Castro}},\ }\bibfield  {title} {\bibinfo {title} {Interaction-driven phases in the half-filled spinless honeycomb lattice from exact diagonalization},\ }\href {https://doi.org/10.1103/PhysRevB.88.245123} {\bibfield  {journal} {\bibinfo  {journal} {Phys. Rev. B}\ }\textbf {\bibinfo {volume} {88}},\ \bibinfo {pages} {245123} (\bibinfo {year} {2013})}\BibitemShut {NoStop}%
\bibitem [{\citenamefont {Jia}\ \emph {et~al.}(2013)\citenamefont {Jia}, \citenamefont {Guo}, \citenamefont {Chen}, \citenamefont {Shen},\ and\ \citenamefont {Feng}}]{PhysRevB.88.075101}%
  \BibitemOpen
  \bibfield  {author} {\bibinfo {author} {\bibfnamefont {Y.}~\bibnamefont {Jia}}, \bibinfo {author} {\bibfnamefont {H.}~\bibnamefont {Guo}}, \bibinfo {author} {\bibfnamefont {Z.}~\bibnamefont {Chen}}, \bibinfo {author} {\bibfnamefont {S.-Q.}\ \bibnamefont {Shen}},\ and\ \bibinfo {author} {\bibfnamefont {S.}~\bibnamefont {Feng}},\ }\bibfield  {title} {\bibinfo {title} {Effect of interactions on two-dimensional dirac fermions},\ }\href {https://doi.org/10.1103/PhysRevB.88.075101} {\bibfield  {journal} {\bibinfo  {journal} {Phys. Rev. B}\ }\textbf {\bibinfo {volume} {88}},\ \bibinfo {pages} {075101} (\bibinfo {year} {2013})}\BibitemShut {NoStop}%
\bibitem [{\citenamefont {Daghofer}\ and\ \citenamefont {Hohenadler}(2014)}]{PhysRevB.89.035103}%
  \BibitemOpen
  \bibfield  {author} {\bibinfo {author} {\bibfnamefont {M.}~\bibnamefont {Daghofer}}\ and\ \bibinfo {author} {\bibfnamefont {M.}~\bibnamefont {Hohenadler}},\ }\bibfield  {title} {\bibinfo {title} {Phases of correlated spinless fermions on the honeycomb lattice},\ }\href {https://doi.org/10.1103/PhysRevB.89.035103} {\bibfield  {journal} {\bibinfo  {journal} {Phys. Rev. B}\ }\textbf {\bibinfo {volume} {89}},\ \bibinfo {pages} {035103} (\bibinfo {year} {2014})}\BibitemShut {NoStop}%
\bibitem [{\citenamefont {Capponi}\ and\ \citenamefont {L\"auchli}(2015)}]{PhysRevB.92.085146}%
  \BibitemOpen
  \bibfield  {author} {\bibinfo {author} {\bibfnamefont {S.}~\bibnamefont {Capponi}}\ and\ \bibinfo {author} {\bibfnamefont {A.~M.}\ \bibnamefont {L\"auchli}},\ }\bibfield  {title} {\bibinfo {title} {Phase diagram of interacting spinless fermions on the honeycomb lattice: A comprehensive exact diagonalization study},\ }\href {https://doi.org/10.1103/PhysRevB.92.085146} {\bibfield  {journal} {\bibinfo  {journal} {Phys. Rev. B}\ }\textbf {\bibinfo {volume} {92}},\ \bibinfo {pages} {085146} (\bibinfo {year} {2015})}\BibitemShut {NoStop}%
\bibitem [{\citenamefont {Motruk}\ \emph {et~al.}(2015)\citenamefont {Motruk}, \citenamefont {Grushin}, \citenamefont {de~Juan},\ and\ \citenamefont {Pollmann}}]{PhysRevB.92.085147}%
  \BibitemOpen
  \bibfield  {author} {\bibinfo {author} {\bibfnamefont {J.}~\bibnamefont {Motruk}}, \bibinfo {author} {\bibfnamefont {A.~G.}\ \bibnamefont {Grushin}}, \bibinfo {author} {\bibfnamefont {F.}~\bibnamefont {de~Juan}},\ and\ \bibinfo {author} {\bibfnamefont {F.}~\bibnamefont {Pollmann}},\ }\bibfield  {title} {\bibinfo {title} {Interaction-driven phases in the half-filled honeycomb lattice: An infinite density matrix renormalization group study},\ }\href {https://doi.org/10.1103/PhysRevB.92.085147} {\bibfield  {journal} {\bibinfo  {journal} {Phys. Rev. B}\ }\textbf {\bibinfo {volume} {92}},\ \bibinfo {pages} {085147} (\bibinfo {year} {2015})}\BibitemShut {NoStop}%
\bibitem [{\citenamefont {Yu}\ \emph {et~al.}(2023)\citenamefont {Yu}, \citenamefont {Zhang}, \citenamefont {Shen},\ and\ \citenamefont {Deng}}]{YU2023}%
  \BibitemOpen
  \bibfield  {author} {\bibinfo {author} {\bibfnamefont {L.-W.}\ \bibnamefont {Yu}}, \bibinfo {author} {\bibfnamefont {S.-Y.}\ \bibnamefont {Zhang}}, \bibinfo {author} {\bibfnamefont {P.-X.}\ \bibnamefont {Shen}},\ and\ \bibinfo {author} {\bibfnamefont {D.-L.}\ \bibnamefont {Deng}},\ }\bibfield  {title} {\bibinfo {title} {Unsupervised learning of interacting topological phases from experimental observables},\ }\bibfield  {journal} {\bibinfo  {journal} {Fundamental Research}\ }\href {https://doi.org/https://doi.org/10.1016/j.fmre.2022.12.016} {https://doi.org/10.1016/j.fmre.2022.12.016} (\bibinfo {year} {2023})\BibitemShut {NoStop}%
\bibitem [{\citenamefont {Sylju\aa{}sen}\ and\ \citenamefont {Sandvik}(2002)}]{PhysRevE.66.046701}%
  \BibitemOpen
  \bibfield  {author} {\bibinfo {author} {\bibfnamefont {O.~F.}\ \bibnamefont {Sylju\aa{}sen}}\ and\ \bibinfo {author} {\bibfnamefont {A.~W.}\ \bibnamefont {Sandvik}},\ }\bibfield  {title} {\bibinfo {title} {Quantum monte carlo with directed loops},\ }\href {https://doi.org/10.1103/PhysRevE.66.046701} {\bibfield  {journal} {\bibinfo  {journal} {Phys. Rev. E}\ }\textbf {\bibinfo {volume} {66}},\ \bibinfo {pages} {046701} (\bibinfo {year} {2002})}\BibitemShut {NoStop}%
\bibitem [{\citenamefont {Sandvik}(2010)}]{10.1063/1.3518900}%
  \BibitemOpen
  \bibfield  {author} {\bibinfo {author} {\bibfnamefont {A.~W.}\ \bibnamefont {Sandvik}},\ }\bibfield  {title} {\bibinfo {title} {{Computational Studies of Quantum Spin Systems}},\ }\href {https://doi.org/10.1063/1.3518900} {\bibfield  {journal} {\bibinfo  {journal} {AIP Conference Proceedings}\ }\textbf {\bibinfo {volume} {1297}},\ \bibinfo {pages} {135} (\bibinfo {year} {2010})},\ \Eprint {https://arxiv.org/abs/https://pubs.aip.org/aip/acp/article-pdf/1297/1/135/11407753/135\_1\_online.pdf} {https://pubs.aip.org/aip/acp/article-pdf/1297/1/135/11407753/135\_1\_online.pdf} \BibitemShut {NoStop}%
\bibitem [{\citenamefont {Dorneich}\ and\ \citenamefont {Troyer}(2001)}]{PhysRevE.64.066701}%
  \BibitemOpen
  \bibfield  {author} {\bibinfo {author} {\bibfnamefont {A.}~\bibnamefont {Dorneich}}\ and\ \bibinfo {author} {\bibfnamefont {M.}~\bibnamefont {Troyer}},\ }\bibfield  {title} {\bibinfo {title} {Accessing the dynamics of large many-particle systems using the stochastic series expansion},\ }\href {https://doi.org/10.1103/PhysRevE.64.066701} {\bibfield  {journal} {\bibinfo  {journal} {Phys. Rev. E}\ }\textbf {\bibinfo {volume} {64}},\ \bibinfo {pages} {066701} (\bibinfo {year} {2001})}\BibitemShut {NoStop}%
\bibitem [{\citenamefont {Krizhevsky}\ \emph {et~al.}(2012)\citenamefont {Krizhevsky}, \citenamefont {Sutskever},\ and\ \citenamefont {Hinton}}]{NIPS2012_c399862d}%
  \BibitemOpen
  \bibfield  {author} {\bibinfo {author} {\bibfnamefont {A.}~\bibnamefont {Krizhevsky}}, \bibinfo {author} {\bibfnamefont {I.}~\bibnamefont {Sutskever}},\ and\ \bibinfo {author} {\bibfnamefont {G.~E.}\ \bibnamefont {Hinton}},\ }\bibfield  {title} {\bibinfo {title} {Imagenet classification with deep convolutional neural networks},\ }in\ \href {https://proceedings.neurips.cc/paper_files/paper/2012/file/c399862d3b9d6b76c8436e924a68c45b-Paper.pdf} {\emph {\bibinfo {booktitle} {Advances in Neural Information Processing Systems}}},\ Vol.~\bibinfo {volume} {25},\ \bibinfo {editor} {edited by\ \bibinfo {editor} {\bibfnamefont {F.}~\bibnamefont {Pereira}}, \bibinfo {editor} {\bibfnamefont {C.}~\bibnamefont {Burges}}, \bibinfo {editor} {\bibfnamefont {L.}~\bibnamefont {Bottou}},\ and\ \bibinfo {editor} {\bibfnamefont {K.}~\bibnamefont {Weinberger}}}\ (\bibinfo  {publisher} {Curran Associates, Inc.},\ \bibinfo {year} {2012})\BibitemShut {NoStop}%
\bibitem [{\citenamefont {Zeiler}\ and\ \citenamefont {Fergus}(2013)}]{zeiler2013visualizing}%
  \BibitemOpen
  \bibfield  {author} {\bibinfo {author} {\bibfnamefont {M.~D.}\ \bibnamefont {Zeiler}}\ and\ \bibinfo {author} {\bibfnamefont {R.}~\bibnamefont {Fergus}},\ }\href@noop {} {\bibinfo {title} {Visualizing and understanding convolutional networks}} (\bibinfo {year} {2013}),\ \Eprint {https://arxiv.org/abs/1311.2901} {arXiv:1311.2901 [cs.CV]} \BibitemShut {NoStop}%
\bibitem [{\citenamefont {Sermanet}\ \emph {et~al.}(2014)\citenamefont {Sermanet}, \citenamefont {Eigen}, \citenamefont {Zhang}, \citenamefont {Mathieu}, \citenamefont {Fergus},\ and\ \citenamefont {LeCun}}]{sermanet2014overfeat}%
  \BibitemOpen
  \bibfield  {author} {\bibinfo {author} {\bibfnamefont {P.}~\bibnamefont {Sermanet}}, \bibinfo {author} {\bibfnamefont {D.}~\bibnamefont {Eigen}}, \bibinfo {author} {\bibfnamefont {X.}~\bibnamefont {Zhang}}, \bibinfo {author} {\bibfnamefont {M.}~\bibnamefont {Mathieu}}, \bibinfo {author} {\bibfnamefont {R.}~\bibnamefont {Fergus}},\ and\ \bibinfo {author} {\bibfnamefont {Y.}~\bibnamefont {LeCun}},\ }\href@noop {} {\bibinfo {title} {Overfeat: Integrated recognition, localization and detection using convolutional networks}} (\bibinfo {year} {2014}),\ \Eprint {https://arxiv.org/abs/1312.6229} {arXiv:1312.6229 [cs.CV]} \BibitemShut {NoStop}%
\bibitem [{\citenamefont {Simonyan}\ and\ \citenamefont {Zisserman}(2015)}]{simonyan2015deep}%
  \BibitemOpen
  \bibfield  {author} {\bibinfo {author} {\bibfnamefont {K.}~\bibnamefont {Simonyan}}\ and\ \bibinfo {author} {\bibfnamefont {A.}~\bibnamefont {Zisserman}},\ }\href@noop {} {\bibinfo {title} {Very deep convolutional networks for large-scale image recognition}} (\bibinfo {year} {2015}),\ \Eprint {https://arxiv.org/abs/1409.1556} {arXiv:1409.1556 [cs.CV]} \BibitemShut {NoStop}%
\bibitem [{\citenamefont {Kingma}\ and\ \citenamefont {Ba}(2017)}]{kingma2017adam}%
  \BibitemOpen
  \bibfield  {author} {\bibinfo {author} {\bibfnamefont {D.~P.}\ \bibnamefont {Kingma}}\ and\ \bibinfo {author} {\bibfnamefont {J.}~\bibnamefont {Ba}},\ }\href@noop {} {\bibinfo {title} {Adam: A method for stochastic optimization}} (\bibinfo {year} {2017}),\ \Eprint {https://arxiv.org/abs/1412.6980} {arXiv:1412.6980 [cs.LG]} \BibitemShut {NoStop}%
\bibitem [{\citenamefont {Chollet}\ \emph {et~al.}(2015)\citenamefont {Chollet} \emph {et~al.}}]{chollet2015keras}%
  \BibitemOpen
  \bibfield  {author} {\bibinfo {author} {\bibfnamefont {F.}~\bibnamefont {Chollet}} \emph {et~al.},\ }\href@noop {} {\bibinfo {title} {Keras}},\ \bibinfo {howpublished} {\url{https://keras.io}} (\bibinfo {year} {2015})\BibitemShut {NoStop}%
\bibitem [{\citenamefont {Arnold}\ and\ \citenamefont {Sch\"afer}(2022)}]{PhysRevX.12.031044}%
  \BibitemOpen
  \bibfield  {author} {\bibinfo {author} {\bibfnamefont {J.}~\bibnamefont {Arnold}}\ and\ \bibinfo {author} {\bibfnamefont {F.}~\bibnamefont {Sch\"afer}},\ }\bibfield  {title} {\bibinfo {title} {Replacing neural networks by optimal analytical predictors for the detection of phase transitions},\ }\href {https://doi.org/10.1103/PhysRevX.12.031044} {\bibfield  {journal} {\bibinfo  {journal} {Phys. Rev. X}\ }\textbf {\bibinfo {volume} {12}},\ \bibinfo {pages} {031044} (\bibinfo {year} {2022})}\BibitemShut {NoStop}%
\end{thebibliography}
\providecommand{\noopsort}[1]{}\providecommand{\singleletter}[1]{#1}%

\end{document}